\documentclass{pas}

\usepackage{multirow}
\usepackage{gensymb}
\usepackage{graphicx}
\usepackage{placeins}
\usepackage{float}

\begin{document}

\lefttitle{Publications of the Astronomical Society of Australia}
\righttitle{Jan K\'{a}ra et al.}

\jnlPage{1}{4}
\jnlDoiYr{2026}
\doival{10.1017/pasa.xxxx.xx}

\articletitt{Research Paper}

\title{A Distinct Type of Standstill in Ultra-Compact White Dwarf Binaries}

\author{\gn{Jan} \sn{K\'{a}ra},$^{1}$ 
        \gn{Liliana} \sn{Rivera Sandoval},$^{1,2}$ 
        \gn{Iwona} \sn{Kotko},$^{3}$ 
        \gn{Jean-Marie} \sn{Hameury},$^{4}$ 
        \gn{Luis E.} \sn{Salazar Manzano},$^{5}$ and
        \gn{Ryan J.} \sn{Oelkers}$^{1,2}$
        }

\affil{
        $^1$Department of Physics and Astronomy, University of Texas Rio Grande Valley, Brownsville, TX 78520, USA\\
        $^2$South Texas Space Science Institute, University of Texas Rio Grande Valley, Brownsville, TX 78520, USA\\
        $^3$Astronomical Observatory, Jagiellonian University, ulica Orla 171, 30-244 Krak\'{o}w, Poland\\
        $^4$Universit\'{e} de Strasbourg, CNRS, Observatoire astronomique de Strasbourg, UMR 7550, 67000 Strasbourg, France\\
        $^5$Department of Astronomy, University of Michigan, Ann Arbor, MI 48109, USA\\
        }

\corresp{J. K\'{a}ra, Email: jan.kara@utrgv.edu}

\citeauth{ }

\history{(Received xx xx xxxx; revised xx xx xxxx; accepted xx xx xxxx)}

\begin{abstract}
AM~CVn stars are ultra-compact, semi-detached binaries consisting of a white dwarf primary and a hydrogen-poor secondary. In this work, we analyse the AM~CVn-type systems MGAB-V270, ZTF18aaxuusk and ASASSN-19ct, which have similar properties  and show unusual post-superoutburst behaviour. We focus on the occurrence of standstill-like phases. In MGAB-V270, superoutbursts are followed by a series of echo outbursts that evolves into a standstill-like phase before the system returns to quiescence. ZTF18aaxuus shows similar post-superoutburst activity, with one echo-outburst series also developing into a standstill-like phase. ASASSN-19ct exhibits comparable echo outbursts but no standstill-like phases. Periodicity analysis using TESS data shows that superhumps can persist during both echo outbursts and standstill-like phases. Although standstills have previously been observed in AM~CVn stars, the standstill-like phases we observe appear to represent a distinct phenomenon associated with post-superoutburst activity. We compare our results to other AM~CVns that may exhibit similar behaviour.
\end{abstract}


\maketitle

\section{Introduction}

Accreting white dwarf binaries are semi-detached binary systems with a white dwarf as the primary star. The secondary star, also called a donor, is filling its Roche lobe which allows for transfer of matter between the two components through the first Lagrangian point $\mathrm{L}_1$. When no strong magnetic field is present, the transferred matter accretes onto the white dwarf via an accretion disc. Depending on the mass transfer rate, the accretion disc can be in a stable low-temperature state, a stable high-temperature state, or it can be in an unstable state during which it alternates between the low and high-temperature states. These changes to high-temperature state typically last for several days and they result in events called dwarf novae (DNe) outbursts, during which the brightness of the system increases by several magnitudes. 
A detailed review of accreting white dwarfs was presented by \cite{2003cvs..book.....W}.

The largest group of accreting white dwarfs are cataclysmic variables (CVs). These systems typically have a main-sequence-like star of a late spectral type as their donor and their orbital periods are between $\sim 80\,\mathrm{min}$ and $\sim 10\,\mathrm{h}$. 
Another group are AM~CVn-type stars, which are closely related to CVs. AM~CVns have orbital periods between $5$ and $70\,\mathrm{min}$ and they have hydrogen-poor donors, which can be either another white dwarf or a semi-degenerate star. As a result of the donors being hydrogen-poor stars, the spectra of AM~CVns are devoid of optical hydrogen lines. Both of these groups show a large variety of outbursting activity which differs in durations, amplitudes, and light curve shape. 
The DNe outbursts observed in the CVs are well explained by the disc instability model \citep[DIM; see][for reviews]{2001NewAR..45..449L, 2020AdSpR..66.1004H} and the same model has been successfully applied to explain the properties of outbursts observed in AM~CVns. 
Systems which show outbursting activity have their mass transfer rates ($\dot{M}_\mathrm{tr}$) between the lower and upper critical values $\dot{M}_\mathrm{crit}^-$ and $\dot{M}_\mathrm{crit}^+$, respectively. This allows for the alternation between low and high-temperature states, which results in the change of ionisation level of the dominant chemical element in the disc and its opacity as well. Systems with $\dot{M}_\mathrm{tr}$ above the critical value $\dot{M}_\mathrm{crit}^+$ at the outer disc radius have their accretion disc in a stable high-temperature state and they typically do not show outbursting activity. Similarly, systems with $\dot{M}_\mathrm{tr}$ below the critical value $\dot{M}_\mathrm{crit}^-$ at the inner disc radius do not exhibit outbursts as their discs are in stable low-temperature states. 
\cite{2012A&A...544A..13K} showed that the DIM can be used to explain the various outbursts shapes of AM~CVns when enhanced mass transfer is included in the model. The enhanced mass transfer rate acts as the determining factor of the outbursts shapes.

The different types of outbursting activity were first detected in DNe and they were used to establish different subtypes of DNe. However, as was shown by \cite{2021MNRAS.502.4953D}, AM~CVns show outbursting variety as diverse as is observed in DNe. The most typical outburst type observed in accreting white dwarfs is a dwarf nova outburst, also called normal outburst. The light curves of normal outbursts have a peak-like shape and they are separated by quiescent phases, during which the accretion disc stays in its low-temperature state. 
Some systems show two distinct types of outbursts - normal outbursts and more energetic superoutbursts; these are the SU UMa systems for DNe, all of which have short orbital periods (less than two hours). Superoutbursts have larger amplitudes than normal outbursts, they last longer, and occur less frequently. Systems in superoutbursts typically show superhumps, which is a brightness variation caused by precession of an elliptical accretion disc, as tidal effects are important in those very compact systems. In some cases, superoutbursts are not followed by a return to the quiescent brightness, but they exhibit a series of rebrightenings during which the accretion disc undergoes several additional outbursts which follow one another. These outbursts are also called echo outbursts and they can be observed in WZ~Sge-type of DNe (a subclass of SU UMas that exhibit only superoutbursts) as well as in AM~CVns \citep[e.g.][]{2021MNRAS.508.3275P}. Some systems have standstills, which are phases of constant increased brightness during which the accretion disc is in a stable, hot-temperature state. This feature is the defining property of Z~Cam-type CVs \citep{2014JAVSO..42..177S}, which show phases of normal outbursts alternating with standstills. The outbursts of Z~Cam-type CVs are typically separated by short quiescent phases, and their peak brightness is larger than their brightness during standstills. Standstills have been also reported in several AM~CVns \citep{2023arXiv230204454K, 2023arXiv230211148K, 2023arXiv230708954K}.

As AM~CVns are more compact systems than CVs, the durations of their outbursts and superoutbursts are usually shorter that what are the typical values for CVs \citep[e.g.][]{2026PASA...43...52K}. This can also complicate analysis of their outbursting activity when only ground-based photometry is available, as these observation lack the necessary time-sampling to capture the brightness variations, as was shown by \cite{2021MNRAS.508.3275P}. Therefore, space-based observatories, such as the Transiting Exoplanet Survey Satellite (TESS), which provide continuous observations spanning long time intervals offer a unique opportunity to study outbursting activity of these systems in detail.

In this paper we present the analysis of three targets, MGAB-V270, ZTF18aaxuusk, and ASASSN-19ct, which were previously classified as outbursting AM~CVn stars \citep{2021AJ....162..113V, 2023MSThesis..SM, vsn-a-22996, vsn-a-23109}. We focus on the study of these targets because previous findings show they have almost identical orbital periods $P_\mathrm{orb}\sim31\,\mathrm{min}$ and because they exhibit similar outbursting behaviour during their post-superoutburst phase. This behaviour includes occurrence of standstill-like phases which appear to be distinct from previously reported standstills observed in AM~CVns. We compare our findings with CR~Boo, V803 Cen, and ASASSN-14cc, which were reported to show long-lasting high states \citep{2021MNRAS.502.4953D}, and with KL~Dra, which shows complex behaviour during its post-superoutburst activity \citep{2026PASA...43...70S}.

\begin{table*}[htb!]
    \centering
    \caption{List of targets analysed in this study. Coordinates and magnitudes G are taken from the Gaia DR3 catalogue \citep{2023A&A...674A...1G}, distances correspond to geometric distances determined from Gaia EDR3 catalogue by  \cite{2021AJ....161..147B}. Orbital periods and superhump period are marked by (orb) and (sh), respectively, $T_\mathrm{SO}$ gives the average superoutburst recurrence times.}
    \begin{tabular}{l c c c c c c}
    \toprule
Name         & RA          & DEC          & Period [min] & G      & Distance & $T_\mathrm{SO}$ [d] \\
\hline
MGAB-V270    & 21:28:22.20 & +63:25:57.26 & $   30.94$(sh)$^{(a)}$ & $20.0$ & $782_{-101}^{+189}$ & 150$^{(g)}$\\
ZTF18aaxuusk & 17:36:17.94 & +75:21:22.82 & $   30.92$(sh)$^{(a)}$ & $20.7$ & $1051_{-383}^{+697}$ & 230$^{(g)}$\\
ASASSN-19ct  & 11:33:15.37 & -37:10:20.36 & $   30.94$(sh)$^{(b)}$ & $17.4$ & $238_{-5}^{+5}$ & 190$^{(g)}$\\
\hline
CR Boo       & 13:48:55.18 & +07:57:35.93 & $ 24.5$(orb)$^{(c)}$ & $15.5$ & $350_{-5}^{+4}$ & 50$^{(g)}$\\
V803 Cen     & 13:23:44.53 & -41:44:29.26 & $ 26.61$(orb)$^{(d)}$ & $15.7$ & $284_{-5}^{+6}$ & 70$^{(g)}$\\
ASASSN-14cc  & 21:39:48.22 & -59:59:33.46 & $ 22.5$(sh)$^{(e)}$ & $18.1$ & $1115_{-64}^{+85}$ & 130$^{(g)}$ \\
KL Dra       & 19:24:38.24 & +59:41:46.43 & $ 25.0$(orb)$^{(f)}$ & $19.1$ & $908_{-79}^{+77}$ & 60$^{(h)}$
    \botrule
    \end{tabular}
    \begin{tabnote}
        $^{(a)}$ \cite{2023MSThesis..SM}; 
        $^{(b)}$ \cite{2026PASA...43...52K};
        $^{(c)}$ \cite{1997PASP..109.1100P};
        $^{(d)}$ \cite{2007MNRAS.379..176R};\\
        $^{(e)}$ \cite{2015PASJ...67L...2K};
        $^{(f)}$ \cite{2002MNRAS.334...87W};
        $^{(g)}$ this work;
        $^{(h)}$ \cite{2026PASA...43...70S};
    \end{tabnote}
    \label{T:list_01}
\end{table*}

\section{Data}

\subsection{TESS photometry}

All three targets were observed by TESS \citep{2015JATIS...1a4003R} throughout its mission. Table~\ref{T:list_01} summarises their basic properties and Table~\ref{tab:TESS_ALL} lists TESS photometry analysed in this paper. 
The available photometry for each sector typically has a baseline near 27 days.
When available, we obtained the $120\,\mathrm{s}$ cadence light curve files using the {\sc Python3} package {\sc Lightkurve} \citep{2018ascl.soft12013L}. These light curve files  provide Simple Aperture Photometry (SAP) flux of the targets, as well as Pre-search Data Conditioning (PDC) corrected SAP flux. PDC is meant to correct for instrumental effects present in the SAP light curve. However, as this method can perform poorly on light curves showing outbursts, we used only SAP flux in our study. 
When the $120\,\mathrm{s}$ cadence was not available, we downloaded cutouts of the full frame images (FFIs) using the {\sc Lightkurve} package and we extracted the light curves using aperture photometry. These light curves provide cadence of $30\,\mathrm{min}$ for sectors 1 - 26,  $10\,\mathrm{min}$ for sectors 27 - 52, an $200\,\mathrm{s}$ for all subsequent sectors. 

All TESS light curves used in this study are presented in Figures \ref{fig:TESS_ALL_MGAB-V270} - \ref{fig:TESS_ALL_ZTF18aaxuusk_b}.

\subsection{Archival photometry}

\begin{figure*}[t!]
    \includegraphics[width=0.99\linewidth]{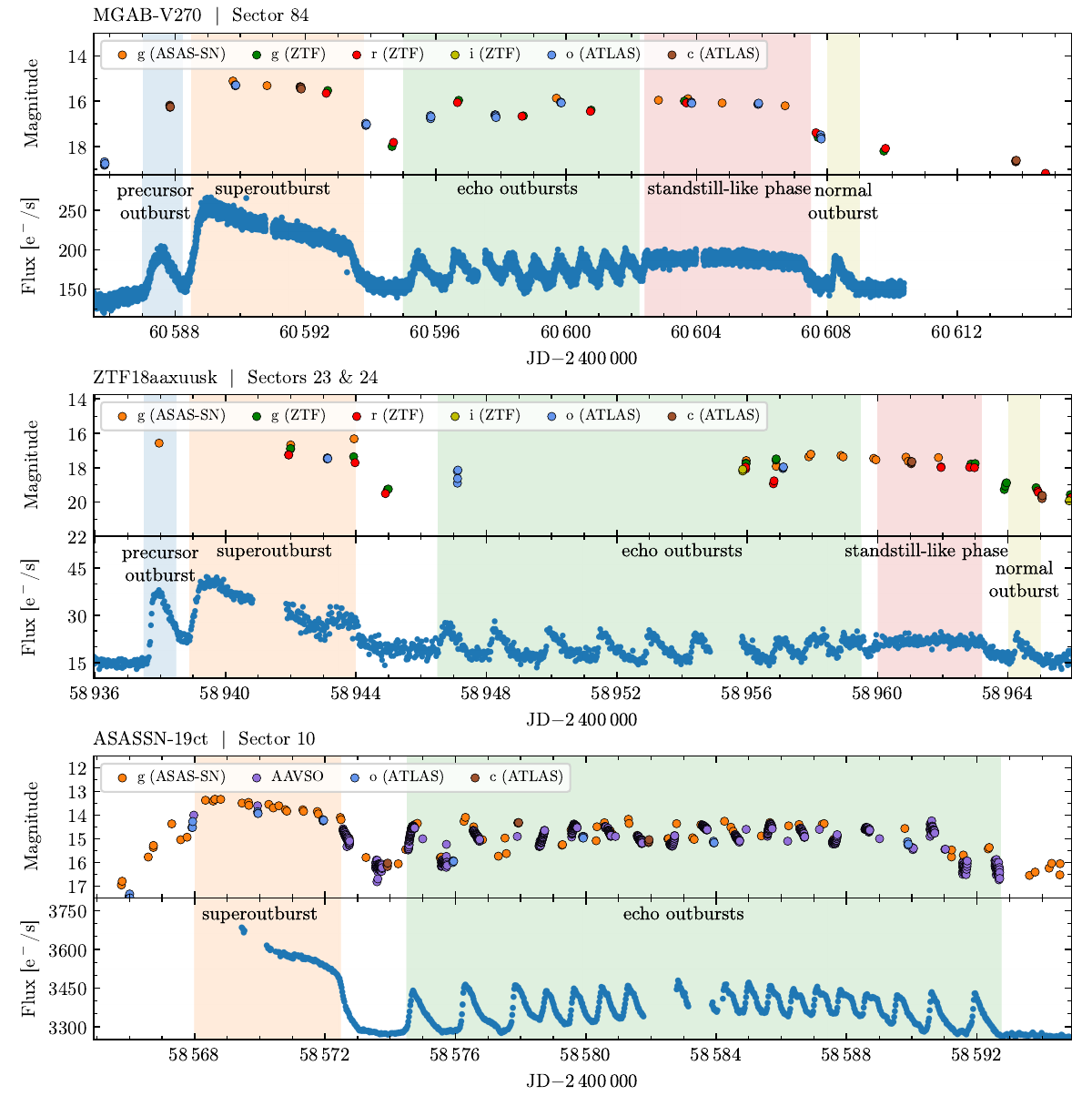}

    \caption{Examples of superoutbursts and subsequent activity of MGAB V270, ZTF18aaxuusk, and ASASSN-19ct as observed by various ground-based observatories and by TESS. Different types of outbursting activity are marked and labelled, all panels show a 30 day span.}
    \label{fig:TRINITY_compare}
\end{figure*}

When available we obtained archival photometry from the Zwicky Transient Facility \citep[ZTF;][]{2019PASP..131g8001G}, the All-Sky Automated Survey for Supernovae \citep[ASAS-SN;][]{2014ApJ...788...48S, 2023arXiv230403791H}, the American Association of Variable Star Observers (AAVSO), the Asteroid Terrestrial-impact Last Alert System \citep[ATLAS;][]{2018PASP..130f4505T, 2020PASP..132h5002S, 2021TNSAN...7....1S}, the Catalina Real-Time Transient Survey \citep[CRTS;][]{2009ApJ...696..870D}, and the Gaia survey \citep{2016A&A...595A...1G, 2023A&A...674A...1G}. 
Light curves composed of this photometry are presented in Figure~\ref{F:LTLC}, comparison of archival photometry and TESS light curve showing outbursting activity are presented in Figures \ref{F:MGAB_COMP_AP} - \ref{fig:19ct_V803_COMP_AP}. Examples of this comparison for the three systems from our study are shown in Figure~\ref{fig:TRINITY_compare} which also marks the different activity phases.

\subsection{Spectroscopic data}

All three targets have available spectroscopic observations. MGAB-V270 (also known as ZTF18abihypg) was spectroscopically analysed by \cite{2021AJ....162..113V}, who obtained a spectrum with the Auxiliary-Port Camera (ACAM) mounted on the 4.2-m William Herschel Telescope at La Palma Observatory. The spectrum has resolution $R \sim 400$ and covers wavelength range $4\,000-9\,000\,\mathrm{\text{\AA}}$.

Spectra of ZTF18aaxuusk and ASASSN-19ct were presented by \cite{2026arXiv260921150M}
and \cite{2026PASA...43...52K}, respectively. Spectra of both systems were obtained with the 8.1-m telescopes at the international Gemini Observatory located on Maunakea in Hawai‘i and Cerro Pachón in Chile. The spectra were obtained with GMOS spectrographs equipped with R150 low-resolution grating which provides resolution of $R\sim600$ and coverage of wavelength range $5\,000-10\,000\,\mathrm{\text{\AA}}$.

\section{Methods}

\subsection{Lomb-Scargle periodograms}

We employed the Lomb-Scargle method \citep{1976Ap&SS..39..447L, 1982ApJ...263..835S} implemented in the {\sc Astropy} package \citep{astropy:2013} to analyse periodicity present in the light curves. Specifically, we focused on periodic variations caused by superhumps during superoutbursts and subsequent rebrightenings. The Lomb-Scargle periodograms were computed using the light curves where long-term trends were subtracted in order to amplify variations occurring at shorter periods. We computed the periodograms for parts of light curves showing specific types of activity, e.g. echo outburst phase.

To evaluate the significance of the detected period, we computed the power level corresponding to $10\%$ false alarm probability (FAP). The FAP level was computed for each individual periodogram using the bootstrap method \citep{2018ApJS..236...16V} implemented in the {\sc Astropy} package. This method estimates the power level of a given probability for a specific temporal grid of observational times in a specified period range. All levels presented in this work were calculated for the period ranges covered by the individual periodograms.

However, this method does not take into account the prior probability of previously detected superhump periods. Therefore, in cases when we detected a peak coinciding with the expected superhump period, we estimated the probability of the periodogram exhibiting its strongest peak within a specified period range when only Gaussian noise is assumed for the flux. To do so, we calculated $50\,000$ Lomb Scargle periodograms of simulated light curves which used the same temporal grid as the observed light curves and whose flux was a Gaussian noise with the same standard deviation as the flux of the de-trended light curves.

To visualise the evolution of periodic variations throughout the different outbursting phases, we computed two-dimensional Lomb-Scargle periodograms of the short-cadence TESS light curves. Each periodogram was constructed by dividing the light curve into bins with length of $1$ day and by computing a Lomb-Scargle periodogram for each individual bin. 
We used the two-dimensional periodograms to select intervals showing periodic variations and we computed one-dimensional Lomb-Scargle periodograms for these intervals.

An example of a two-dimensional periodogram is shown in Figure~\ref{fig:2D_MGABV270_S76} and all two-dimensional periodograms used in this study are presented in Figures~\ref{fig:2D_LomScargle_all} and \ref{fig:2D_LomScargle_V803}. Figure~\ref{fig:MGABV270_1dLS_d76} shows an example of Lomb-Scargle periodograms of specific activity phases of MGAB V270, all periodograms of specific activity phase are presented in Figures~\ref{fig:MGABV270_1dLS_all} and \ref{fig:ztf_803_kl_1dLS_14cc}.

\subsection{Long-term variability}

We used the available ground-based photometry to compile long-term light curves which we used to study the long-term variability of the systems and to compare TESS photometry with the ground-based observations. We used superoutbursts captured by the light curves to determine the average superoutburst recurrence times ($T_{\rm SO}$) listed in Table~\ref{T:list_01} as the average value of time elapsed between the maxima of consecutive superoutbursts.

\section{Results}

\subsection{MGAB-V270}

MGAB-V270 was discovered as a transient by \cite{2020AJ....159..198S} who classified the system as a CV candidate. 
It was spectroscopically classified as an AM CVn-type star by \cite{2021AJ....162..113V} who 
used its superoutburst recurrence time $T_\mathrm{SO}\sim160\,\mathrm{d}$ to estimate its orbital period to be $P_\mathrm{orb}\sim31.2\,\mathrm{min}$. \cite{2021MNRAS.508.3275P} reported on two normal outburst which were observed by TESS, both of them lasted for about one day. 
\cite{2023MSThesis..SM} used subsequent TESS observations showing superoutburst to determine the superhump period of this binary $P_\mathrm{sh}=30.94\,\mathrm{min}$.

This system exhibits both normal outbursts and superoutbursts, the latter of which are typically followed by a series of short echo outbursts, each of which lasts about one day. While ground-based surveys detected numerous superoutburst events, their analysis is limited by the poor coverage of about one observation per night. The TESS observations show superoutbursts and subsequent series of echo outbursts in three different sectors: 58, 76, and 84. All three light curves are presented in Figure~\ref{fig:TESS_ALL_MGAB-V270}. A comparison between ground-based photometry and the TESS light curves is shown in Figure~\ref{F:MGAB_COMP_AP}. While the light curve of Sector 58 does not cover the full echo outburst series, the other two light curves captured the whole event. In all sectors, all the echo outbursts have almost the same maximum brightness, however, the minimum brightness between individual echo outbursts exhibits an increasing trend in all three sectors and consistently remains higher than the system's average quiescent brightness.
The light curves from Sectors 76 and 84 show a series of echo outbursts which transition into a standstill-like state during which the system's brightness is equal to the peak brightness of echo outbursts. These standstill-like states last about 5 days, after which the system's brightness decreases. Sector 76 shows two additional echo outbursts following the standstill-like state, while Sector 84 shows only one additional echo outburst. The system then returns to quiescence. It is possible that the system exhibited the standstill-like phase also during sector 58, however, it cannot be confirmed due to the incomplete coverage of the TESS photometry and insufficient sampling of the ground based data. 

\begin{figure}
    \centering
    \includegraphics[width=0.99\linewidth]{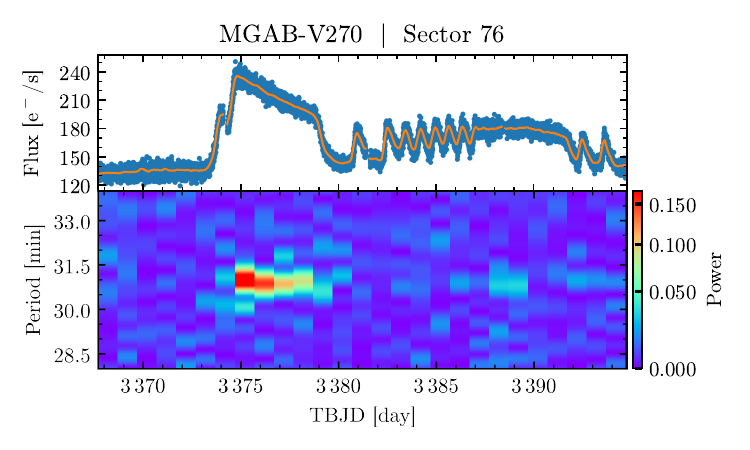}
    \caption{Two-dimensional Lomb-Scargle periodogram of MGAB-V270 TESS light curve from Sector 76. Top panel shows the TESS photometry (blue points) and average light curve (orange line). 
    Bottom panel shows two-dimensional Lomb-Scargle periodogram based on the presented light curve.}
    \label{fig:2D_MGABV270_S76}
\end{figure}

Figure~\ref{fig:2D_LomScargle_all} shows two-dimensional Lomb-Scargle periodograms of the three TESS light curves showcasing outbursting activity (periodogram for Sector 76 is presented also in Figure~\ref{fig:2D_MGABV270_S76}). 
All three light curves show periodic variations corresponding to the superhump period $P_\mathrm{sh}\doteq31\,\mathrm{min}$ of the system clearly visible during the superoutburst. Additionally, Sector 58 light curve exhibits a variation with similar period during the echo outbursts. Figure~\ref{fig:MGABV270_1dLS_all} shows a periodogram of the whole echo-outburst phase observed in Sector 58 as well as the specific part showing the periodic variations and compares it to the superoutburst periodogram. All three periodograms show peaks at the same period, however, the peaks in the echo outbursts periodograms show low significance with their peak power corresponding to $\mathrm{FAP}\sim70\%$. 
Nevertheless, as the probability of Gaussian noise producing a periodogram exhibiting its strongest peak within $0.1\,\mathrm{min}$ of the detected superhump period is only $\sim1\%$, the detected peak likely correspond to a weak brightness modulation caused by superhumps. 

\begin{figure}
    \centering
    \includegraphics[width=0.99\linewidth]{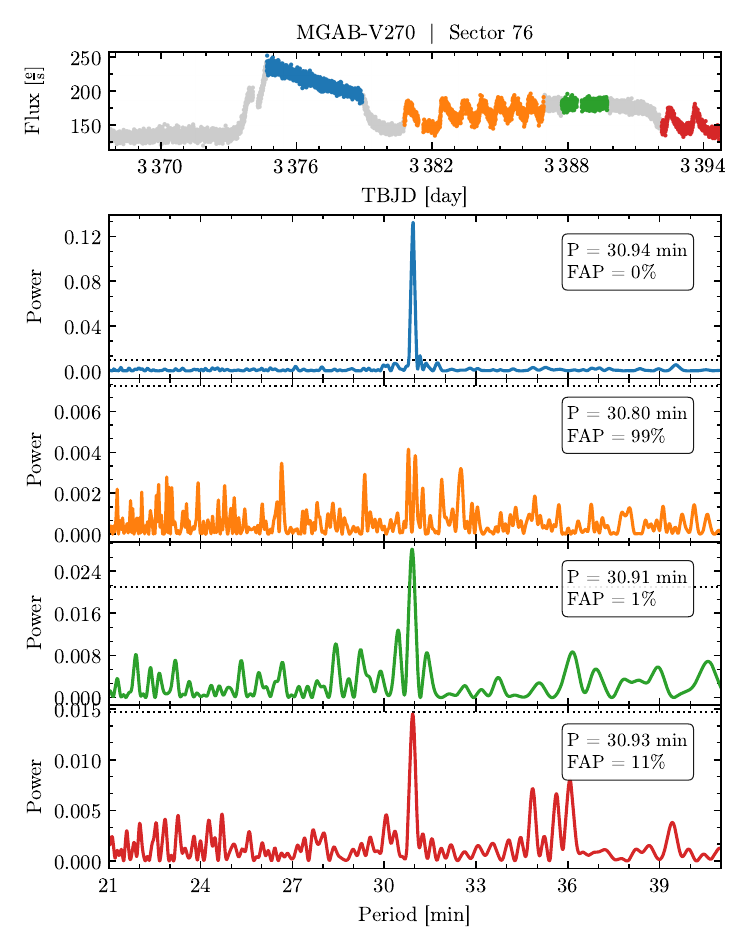}
    \caption{Lomb-Scargle periodograms of different activity phases of MGAB-V270 based on TESS light curve from Sector 76. 
    The dotted line marks the power level corresponding to a $10\%$ FAP, the period and its FAP value of the strongest peak  within $30\,\mathrm{s}$ of the superhump period is presented for each periodogram.
    Note that both orange and green sections of the light curve contain a short observational gap.
    }
    \label{fig:MGABV270_1dLS_d76}
\end{figure}

\begin{figure}
    \centering
    \includegraphics[width=0.99\linewidth]{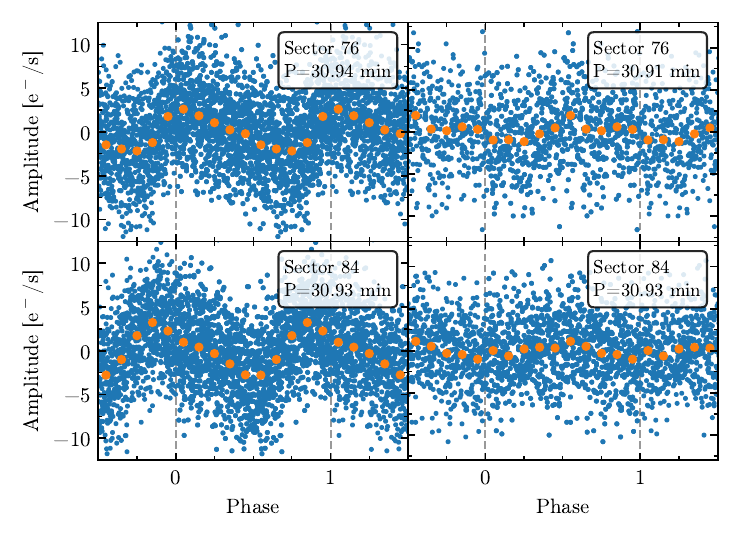}
    \caption{Phase-folded light curves of plateaus (left panels) and standstill like phases (right panels) of MGAB-V270 showing periodic variations. Top-right panel shows photometry obtained between TBJD $3\,387.70$ and $3\,389.71$, bottom-right panel shows photometry obtained between TBJD $3\,604.60$ and $3\,607.13$. Orange points show binned light curves with 10 bins per phase.}
    \label{fig:MGABV270_stand_phase}
\end{figure}

The light curves of Sectors 76 and 84 do not show any significant periodic variations during the echo outbursts, as can be seen also in periodograms shown in Figure~\ref{fig:MGABV270_1dLS_all} (periodogram for Sector 76 is presented also in Figure~\ref{fig:MGABV270_1dLS_d76}). However, they show significant periodic variations occurring during the standstill-like phases. They do not appear throughout the whole standstill-like phase and their peak power in periodograms corresponds to $\mathrm{FAP}\sim1\%$. As the period of these variation matches with the superhump period observed during the superoutbursts, we interpret these variations as superhump modulation of the light curve. Figure~\ref{fig:MGABV270_stand_phase} shows the phase-folded light curves of the plateau phases of the superoutburst and standstill-like phases. The amplitude of modulation is clearly smaller during the standstill-like phase, with the binned light curve varying by about $1\%-2\%$, while the plateau phase exhibits variation of $2\%-3\%$.

The light curve of Sector 76 shows also periodic variation occurring during the two normal outbursts following the standstill-like phase, its period matches that of superhump variations. While the FAP of the detected peak is only $11\%$, the probability of a prominent peak occurring within $0.1\,\mathrm{min}$ of the superhump period is only $\sim1\%$. Therefore, we interpret this signal as superhump variations.

\subsection{ZTF18aaxuusk}

ZTF18aaxuusk 
was discovered by \cite{2014ATel.6163....1B} who classified it as a CV. \cite{Kojiguchi2020} suggested that it might be an AM~CVn system similar to V803 Cen. The system was analysed by \cite{2023MSThesis..SM}, who used TESS photometry to classify this system as an AM~CVn candidate with a superhump period of $30.92\,\mathrm{min}$. This classification was confirmed by \cite{2026arXiv260921150M}, who presented spectroscopic observations of this system. 

ZTF18aaxuusk was observed by TESS during 41 sectors and showed outbursting activity in 10 sectors. All TESS light curves exhibiting this activity are presented in Figures~\ref{fig:TESS_ALL_ZTF18aaxuusk_a} and \ref{fig:TESS_ALL_ZTF18aaxuusk_b}. Selected examples of TESS light curves showing activity and their comparison with ground-based observations are presented in Figure~\ref{F:ZTF_COMP_AP}.
In total, TESS observed five superoutbursts, all of which were followed by a series of echo outbursts. Additionally, Sector 40 shows a series of echo outbursts which followed a superoutburst not covered by TESS observations. 
All echo outbursts show the same maximal brightness and the same length of about 1 day.
The minimum brightness between individual echo outbursts is typically comparable to the quiescent brightness of the system. However, as ZTF18aaxuusk is a faint object ($\mathrm{G}=20.7$), it is possible that the TESS flux during low brightness stages is dominated by background sources and, therefore, it does not reflect actual brightness behaviour of the system. One of the recorded echo outbursts series shows a standstill-like event similar to those observed in MGAB-V270. This standstill-like event occurred during Sector 24 and lasted for about 3 days. The minimum brightnesses between the echo outbursts preceding the standstill-like events show a similar behaviour to that of MGAB-V270. The light curve of Sector 24 has a cadence of $30\,\mathrm{min}$, which is comparable to the superhump period of this system and, therefore, it is not suitable to test the presence of superhumps.

We analysed the available high-cadence light curves of ZTF18aaxuusk for any periodic variations occurring during the echo outbursts and we detected a strong signal only in Sector 83. Two-dimensional Lomb-Scargle periodograms of this sector is presented in Figure~\ref{fig:2D_LomScargle_all}, periodograms of specific parts of the light curves are shown in Figure~\ref{fig:ztf_803_kl_1dLS_14cc}. In this sector, we detected a strong periodicity occurring during one of the echo outbursts with its period matching the superhump period detected during the preceding superoutburst.

\subsection{ASASSN-19ct}

ASASSN-19ct was discovered by the ASAS-SN survey as a transient source and was classified as an AM~CVn candidate by \cite{vsn-a-22996, vsn-a-23109} based on the superoutburst properties and the detected superhump period $P_\mathrm{sh}=31.2\,\mathrm{min}$. This classification was spectroscopically confirmed by \cite{2026PASA...43...52K} who also presented the available TESS photometry of this target and and used it to determined the superhump period $P_\mathrm{sh}=30.94\,\mathrm{min}$. 

ASASSN-19ct exhibits normal outbursts as well as superoutbursts which are always followed by a series of echo outbursts.
TESS light curves of ASASSN-19ct show outbursting activity in three sectors, all three light curves are presented in Figure~\ref{fig:TESS_ALL_ASASSN-19ct}. 

Sector 10 light curves shows a partially covered superoutburst which was followed by a series of echo outbursts. Each echo outburst lasted about one day and they showed similar peak brightness. While the majority of echo outbursts showed a fast rise followed by a slower decline, some of them showed a short plateau-like phases preceding their decline in brightness. The minimum brightness between echo outbursts gradually increased before returning to the quiescent level, as in the previous two systems. 
This behaviour can be discerned also in the ground-based photometry presented in Figure~\ref{fig:19ct_V803_COMP_AP}. The phase of echo outburst series during which echo outbursts with plateau-like phases occurred is also the phase showing the highest minimum brightness separating the echo outbursts. Cadence of Sector 10 light curve is $30\,\mathrm{min}$, therefore it is not suitable to analyse superhump occurrence.

Another superoutburst was partially detected during Sector 37. This light curve covers onset of this superoutburst and beginning of its plateau phase. It also shows a precursor outburst which occurred about 5 days before the onset of the superoutburst. A two-dimensional periodogram analysing this light curve is shown in Figure~\ref{fig:2D_LomScargle_all}. It clearly shows a presence of superhumps which appeared during the precursor outburst and became strongest during the plateau phase. This early occurrence of superhumps was also reported for this superoutburst by \cite{2023MSThesis..SM}.

The light curve from Sector 63 shows one normal outburst which showed a symmetrical shape. The light curve does not show any significant periodic variations.

While this system does not exhibit a post-superoutburst standstill phase, the properties of its echo outburst phase (notably the evolution of the minimum brightness separating the echo outbursts) are similar to those observed in MGAB-V270 and ZTF18aaxuusk. Therefore, we expect that the mechanism responsible for occurrence of standstill-like phases of MGAB-V270 and ZTF18aaxuusk could be also responsible for the properties of echo outburst phase in ASASSN-19ct. Also, as not all superoutbursts of ZTF18aaxuusk were followed by a standstill-like phase, it is possible that the same might be true for ASASSN-19ct and standstill-like phases might exist in ASASSN-19ct as well.

\subsection{Other AM~CVns with similar behaviour}

In order to compare targets from this study to other known AM~CVns which might exhibit similar behaviour, we obtained available archival photometry of CR~Boo, V803~Cen, and ASASSN-14cc. These three systems were identified by \cite{2021MNRAS.502.4953D} as AM~CVns with unusually long superoutburst, and CR~Boo and V803~Cen were reported by the same authors as systems showing standstills. The standstill of CR Boo was also analysed by \cite{2023arXiv230204454K} who classified this system as a helium analogue of Z Cam-type CV. 
We also included KL~Dra in our comparison. This system is well covered by TESS observations as it lies in its continuous viewing zone and also displays similar standstill behaviour in some supercycles. A detailed analysis of these observations was presented by \cite{2026PASA...43...70S}, who used 41 TESS sectors to analyse the long-term variability of KL Dra.
We also note that \cite{2023arXiv230211148K} reported occurrence of standstills in an AM~CVn star MGAB-V240. While this system shows both superoutbursts and normal outbursts, its superoutbursts are not followed by rebrightenings. Therefore, we did not include this system in our comparison.

\begin{figure}[tb]
    \includegraphics[width=0.99\linewidth]{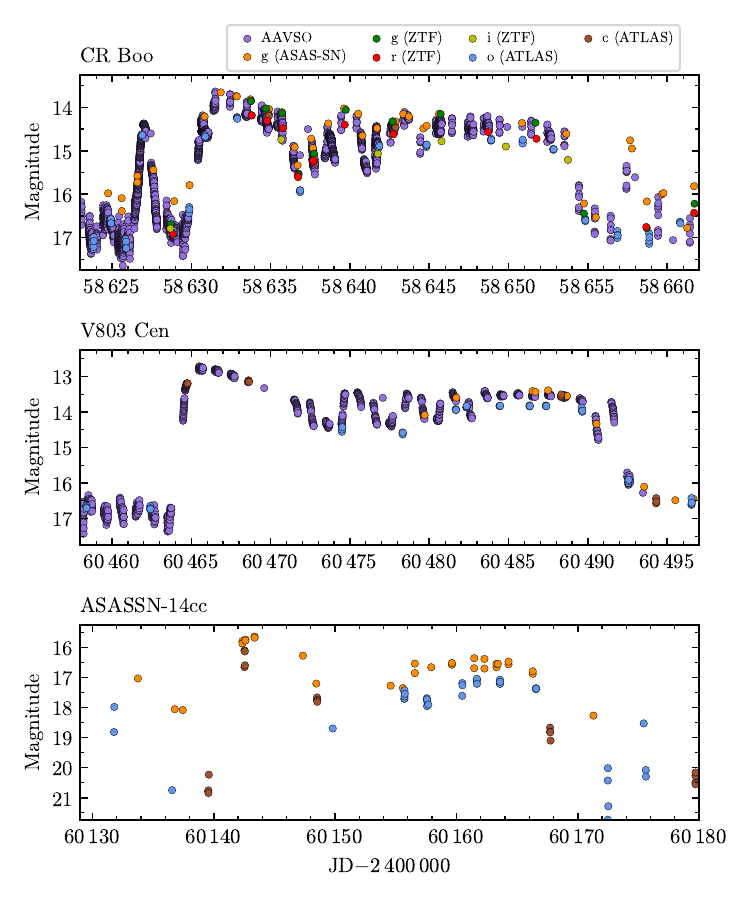}

    \caption{Examples of ground-based photometry of superoutbursts and subsequent brightness behaviour of AM~CVns with reported long superoutbursts.}
    \label{fig:KNOWN_EXAMPLES}
\end{figure}

The available archival photometry of CR Boo, V803 Cen, and ASASSN-14cc is presented in Figure~\ref{F:LTLC}, selected examples of their superoutbursts and subsequent brightness evolution are presented in Figure~\ref{fig:KNOWN_EXAMPLES}. CR Boo and V803 Cen show superoutbursts which are typically followed by rebrightenings and their recurrence times are 50 days and 70 days, respectively. Some of the rebrightenings show behaviour which could be interpreted as a phase of constant brightness similar to the standstill-like phase observed in our findings of MGAB-V270 and ZTF18aaxuusk, but the limited coverage of ground-based photometry does not allow us to determine this unambiguously. Superoutbursts of ASASSN-14cc are not typically followed by rebrightenings. In some cases, the system shows behaviour which could be interpreted as a superoutburst followed by rebrightenings, an example of this is shown in Figure~\ref{fig:KNOWN_EXAMPLES}. However, the observational coverage of these events is too scarce to verify the presence of rebrightenings.

While CR Boo was not observed by TESS, V803 Cen and ASASSN-14cc were observed during multiple TESS sectors. Figures~\ref{fig:TESS_ALL_V803Cen} and \ref{fig:TESS_ALL_ASASSN-14cc} show their TESS light curves which exhibit outbursting activity.

\subsubsection{V803 Cen}

V803 Cen shows large variety in its outbursting activity. It consists of normal outbursts, superoutbursts followed by rebrightenings, and standstills. The latter are typically  accompanied by occurrence of outbursts, which resembles the standstill behaviour of some IW And-type CVs, which are a sub-type of Z Cam CVs \citep{2013PASP..125.1421S, 2019PASJ...71...20K, 2024ApJ...976..107S}. However, standstills of IW And-type CVs are terminated by an outburst, which cannot be explained by the DIM, since the disc is already in a hot state during the standstill \citep{2014A&A...569A..48H}. On the other hand, V803 Cen does not show such behaviour. Its disc luminosity drops after a superoutburst and a sequence of very shortly spaced outbursts may occur. At the end of the short outbursts, the luminosity drops below the luminosity seen at the end of a superoutburst plateau. This can be explained by the DIM, since a cooling front can start (but never reaches the inner edge of the disc), in contrast with IW And-type CVs where the disc appears to be always in a hot state.

TESS light curves show all types of outbursting activity mentioned above, as can be seen in Figure~\ref{fig:TESS_ALL_V803Cen}. The TESS light curve from Sector 101, which was obtained during a high-brightness state, exhibits also a superoutburst of duration about 3 days, which is shorter than the duration of outbursts detected in the Sector 64, which lasted for about 6 days. Figure~\ref{fig:19ct_V803_COMP_AP} shows a comparison to the TESS light curves with ground based photometry. While the Sector 64 light curve does not show the full duration of the rebrightenings, the ground based light curve captured the whole event, even though at lower sampling. The ground based light curves shows a variation corresponding to the echo outbursts detected in the TESS light curve, but it also shows a constant brightness between $\mathrm{JD}\;2\,460\,077$ and $2\,460\,082$. This could be interpreted as a standstill-like phase similar to the ones observed in MGAB-V270 and ZTF18aaxuusk . The luminosity during this phase is about equal to the luminosity at the end of the plateau phase of the superoutburst, just before the onset of the cooling front, which means that the disc can be fully in the hot state, supporting our interpretation. However, we cannot rule out that it is an effect caused by poor sampling.

The orbital period of V803 Cen was determined by \cite{2007MNRAS.379..176R} from spectroscopic observations to be $P_\mathrm{orb}=26.61\pm0.02\,\mathrm{min}$.
This system was recently analysed by \cite{2026ApJ...996..113S} who used available photometry including TESS data from sectors 11, 37 and 64. The authors reported occurrence of positive superhumps which were detected during different outbursting stages, including rebrightenings and the standstill-like phases.

Light curves from Sector 11 has cadence of $30\,\mathrm{min}$ and as such it is not suitable for analysis of orbital and superhump variations. We analysed the light curves from Sectors 37, 64, and 101, their two-dimensional periodograms are presented in Figure~\ref{fig:2D_LomScargle_V803}. Periodograms of specific sections of Sectors 64 and 101 are shown in Figure~\ref{fig:ztf_803_kl_1dLS_14cc}.
Light curve from Sector 37 shows occurrence of two normal outbursts. Its two-dimensional periodogram shows periodic variation with $P=26.8\,\mathrm{min}$ occurring before the onset of the first outburst. This agrees with the findings of \cite{2026ApJ...996..113S} who interpreted this periodicity as superhump variations.
Light curves from Sectors 64 and 101 show a strong periodic signal connected to superhump occurrence during the superoutbursts, the Sector 64 light curve also shows similar variation occurring during the precursor outburst. Both light curves also show periodic variation occurring during the series of consecutive outbursts. This periodicity, however, differs from the one detected during superoutbursts. In Sector 64, we detected period $P_\mathrm{sh}=26.87\pm0.04\,\mathrm{min}$\footnote{The period and its error was determined by fitting the peak in periodogram with a Gaussian profile and adopting its position as the period and its width as the error.} during the superoutbursts and period $P=26.97\pm0.03\,\mathrm{min}$ during the rebrightenings. The latter value agrees within the error with the periods determined for the series of outbursts observed in Sector 101, for which we obtained $P=26.98\pm0.02\,\mathrm{min}$ and $P=26.99\pm0.03\,\mathrm{min}$.

\begin{figure*}
    \centering
    \includegraphics[width=0.49\linewidth]{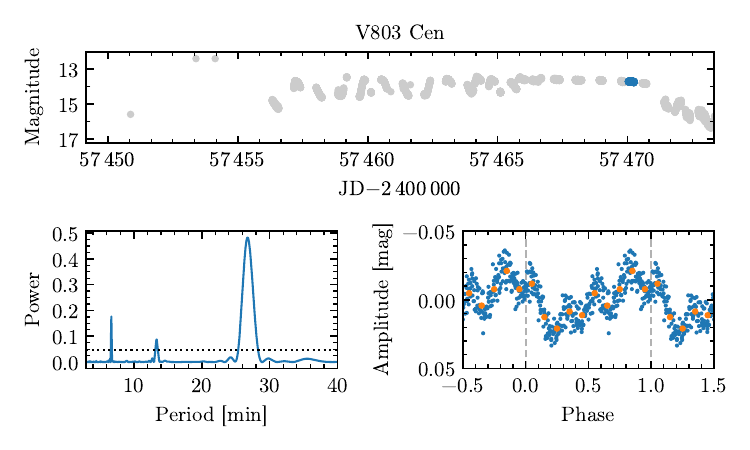}
    \includegraphics[width=0.49\linewidth]{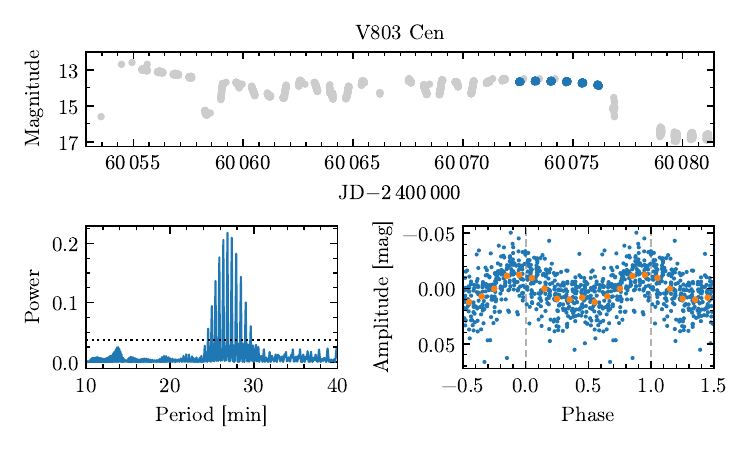}
    \caption{Lomb-Scargle periodograms of AAVSO observations of V803 Cen obtained during a potential standstill-like events (left-bottom panels) and resulting phase-folded light curve (bottom right panels). Top panels shows the sections of light curve used in the analysis. The black dotted line marks the $10\%$ FAP level, the orange points show binned light curve.}
    \label{fig:V803CEN_aavso_period}
\end{figure*}

We also tested the occurrence of superhumps during two potential standstill-like events captured by the AAVSO data. The analysed regions of the light curves, their periodograms and phase-folded light curves are presented in Figure~\ref{fig:V803CEN_aavso_period}. The first region consists of data obtained on $\mathrm{JD}\,2\,457\,474$ which span about 6 hours, the other region consists of observations obtained between $\mathrm{JD}\,2\,460\,077$ and $\mathrm{JD}\,2\,460\,083$. For the analysis, we subtracted $1.2\,\mathrm{h}$ averaged light curves from the observed ones to minimise any long-term components of the periodogram and to reduce the scatter of the phase-folded light curves. The periodograms clearly show a periodic signal, the strongest peak corresponds to a period $P=26.76\pm0.70\,\mathrm{min}$ in the first region and $P=26.9\pm1.6\,\mathrm{min}$\footnote{We fitted the maxima of prominent peaks with a Gaussian function and adopted its width as the uncertainty.} in the other one, FAPs are in both cases $0\%$. These periods agree with the superhump period of this system as well as with the periods detected during the rebrightenings and during the standstills.
While the phase-folded light curve of the second region clearly shows single-humped superhump variations, the light curve of the first region also shows an additional modulation with a period $P=6.69\pm0.04\,\mathrm{min}$ which is four times shorter than the superhump period.

While this system shows strong similarities with the rebrightenings of MGAB-V270, ZTF18aaxuusk, and ASASSN-19ct, it is a more active system. The recurrence time of its superoutbursts is about 70 days, which is shorter than the recurrence times of targets from our study. This high activity of V803 Cen suggests that the mass-transfer rate of this system is higher than the mass-transfer rates of MGAB-V270, ZTF18aaxuusk, or ASASSN-19ct. Considering the fact that V803~Cen also shows long high-brightness states, we expect its mass-transfer rate to be close to the critical value of $\dot{M}_\mathrm{crit}^+$.

\subsubsection{ASASSN-14cc}

TESS light curves of ASASSN-14cc show superoutbursts in Sectors 1, 28, 103, and 104, even though the light curve of superoutburst in Sector 28 is strongly affected by TESS instrumental variations and superoutburst in Sector 103 was covered only partly. Superoutbursts in Sectors 1, 103, and 104 were clearly followed by return to the quiescent level and none of the four superoutbursts observed by TESS show any rebrightenings occurring in the system. Figure~\ref{fig:ztf_803_kl_1dLS_14cc} presents the periodograms of the superoutbursts from Sectors 103 and 104 which show a strong signal corresponding to the superhump period at $P_\mathrm{sh}=22.46\pm0.04\,\mathrm{min}$ and $P_\mathrm{sh}=22.45\pm0.02\,\mathrm{min}$, respectively.
The duration of the fully observed superoutburst is $\sim7$ days, which is much shorter than the length of outbursting activity shown in Figure~\ref{fig:KNOWN_EXAMPLES}, which lasted for about $35$ days. This suggests that the outbursting activity presented in Figure~\ref{fig:KNOWN_EXAMPLES} is not a singular superoutburst, but rather a superoutburst which was followed by rebrightenings. This is also supported by the significant scatter in the observed light curve with low luminosities below that at the end of the superoutburst plateau, which means that the disc cannot longer be fully maintained in the hot state.

\subsubsection{KL Dra}

\begin{figure}
    \centering
    \includegraphics[width=0.99\linewidth]{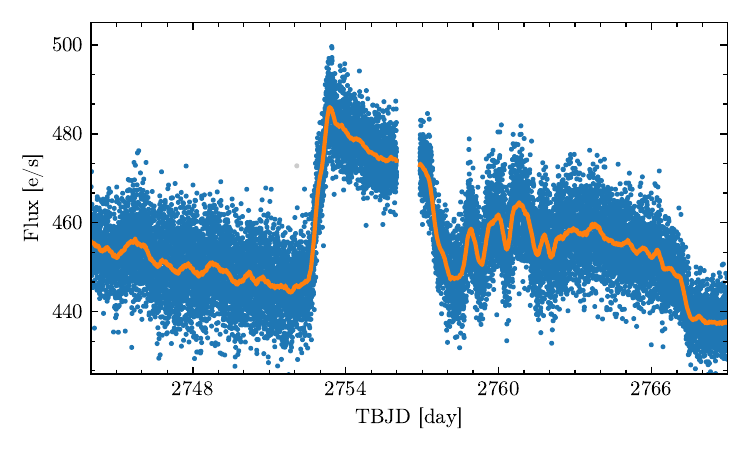}
    \caption{TESS light curve of KL Dra from Sector 53 showing an example superoutburst followed by rebrightenings. The flux of the second half of the light curve were shifted in such way that the plateau phase  shows a continuous brightness evolution.
    }
    \label{fig:KL_Dra_LC}
\end{figure}

KL Dra is an AM~CVn system with orbital period $P_\mathrm{orb}=25\,\mathrm{min}$ \citep{2002MNRAS.334...87W} which exhibits superoutbursts followed by rebrightenings. Example of this behaviour is presented in Figure~\ref{fig:KL_Dra_LC}, for more examples and a detailed analysis of its outbursting behaviour see \cite{2026PASA...43...70S}. The rebrightenings of KL Dra are typically ended by a long phase of high brightness after which the system returns to its quiescence. 
The high-brightness states typically last 3 - 7 days and their average brightness is smaller than the peak brightness of the preceding echo outbursts.
They can exhibit a periodic variation which is likely caused by superhumps, as can be seen in the periodograms presented in Figure~\ref{fig:ztf_803_kl_1dLS_14cc}. 
The detected periodicity corresponds to period $P=25.57\pm0.06\,\mathrm{min}$, which agrees within the error with the superhump period $P_\mathrm{sh}=25.49\pm0.06\,\mathrm{min}$ detected during the preceding superoutburst.
Here we present only one example, a more detailed study of periodic variation in TESS data of KL Dra will be presented by Salazar Manzano et al. (in prep).

The behaviour of KL Dra shares some similarities with the one observed in MGAB-V270 and ZTF18aaxuusk, however, there are also multiple differences. The echo outbursts of KL Dra do not share the same maximal brightness, duration, or their shape. Furthermore, the high-brightness phase shows more complex behaviour and its average brightness is smaller than the maximal brightness of echo outbursts.

\section{Discussion}

\subsection{Spectroscopy \& colour-magnitude diagram}

\begin{figure*}
    \centering
    \includegraphics[width=0.99\linewidth]{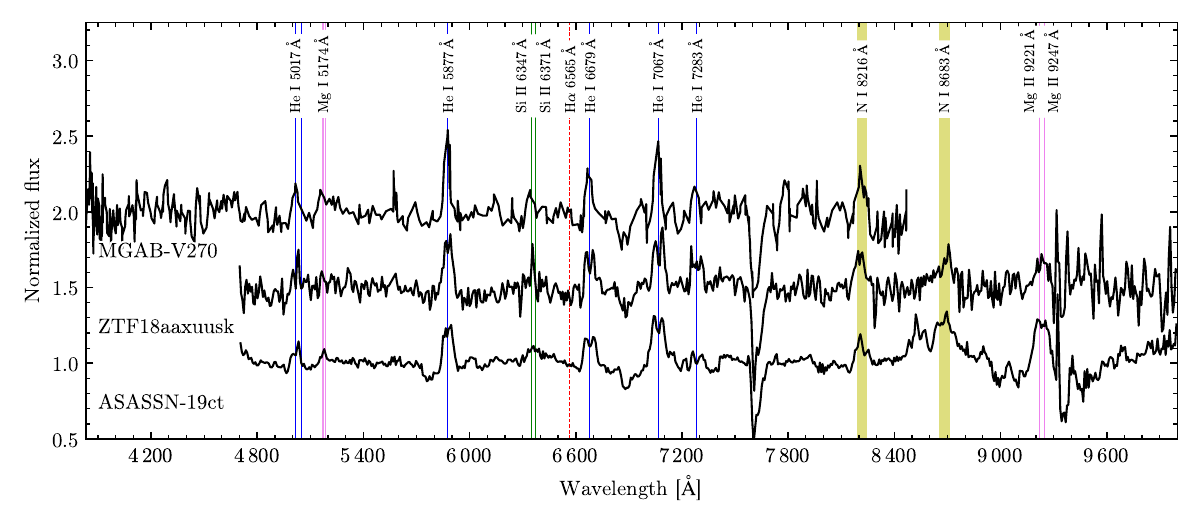}
    \caption{Spectra of MGAB-V270, ZTF18aaxuusk, and ASASSN-19ct. Spectrum of MGAB-V270 is adapted from \cite{2021AJ....162..113V}.
    }
    \label{fig:specrta}
\end{figure*}

\begin{figure}[tbh]
    \centering
    \includegraphics[width=0.99\linewidth]{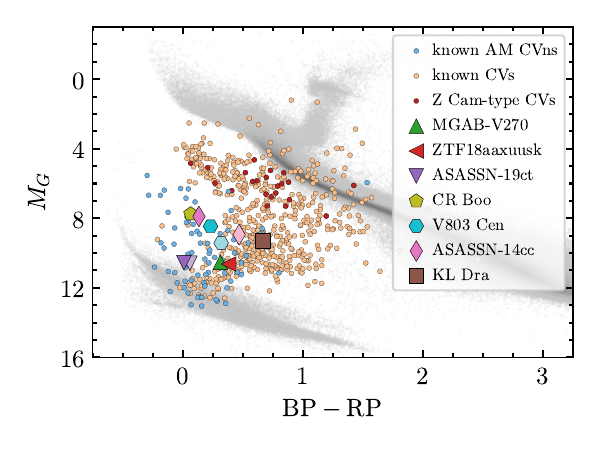}

    \caption{Colour-magnitude diagram based on Gaia photometry. Known AM~CVns and CVs are taken from the catalogues of \cite{2025A&A...700A.107G} and \cite{2003A&A...404..301R}, respectively. Systems studied in this paper are shown as triangles, positions of AM~CVns discussed in this paper are marked by individual symbols. Quiescent properties of targets with available Gaia epoch photometry are marked by corresponding light-coloured symbols.}
    \label{fig:CMD}
\end{figure}

Figure~\ref{fig:specrta} shows the normalised spectra of MGAB-V270, ZTF18aaxuusk, and ASASSN-19ct. Spectra of all three targets were obtained during quiescence. The spectra are devoid of hydrogen lines, as is typical for AM~CVn systems, and they show similar spectral features which suggests that the systems have similar composition and properties. Specifically, it suggests that the donor stars of these system are of a similar type, and therefore the mass transfer rates and its behaviour could be similar as well. 
The emission lines of all three are double peaked, which suggests they originate in an accretion disc observed at high inclination.
The similarity of these systems is illustrated by their position in the colour-magnitude diagram presented in Figure~\ref{fig:CMD}, which shows that all three systems have very similar absolute magnitude $M_\mathrm{G}$. However, while MGAB-V270 and ZTF18aaxuusk also show very similar $\mathrm{BP}-\mathrm{RP}$ colours, ASASSN-19ct appears to be bluer.

The optical spectrum of CR Boo was analysed by \cite{1987ApJ...313..757W} who reported shallow helium absorption in the system's high state and weak emission feature during its low state, both of these cases are different from the spectra of targets presented in Figure~\ref{fig:specrta}. Similar properties were found in the spectra of V803 Cen by \cite{1987MNRAS.227..347O, 1989MNRAS.236..319O}, who reported only weak emission features appearing during the low state. Optical spectra of V803 Cen were also analysed by \cite{2007MNRAS.379..176R}, who observed several helium emission lines during the system's low state, all of which were double peaked. 
The optical spectra of KL Dra analysed by \cite{2010MNRAS.407.1819R} show double-peaked helium emission lines in low state and broad helium absorption lines in high state together with helium emission line at $4\,686\,\mathrm{\text{\AA}}$.

While the $\mathrm{BP}-\mathrm{RP}$ colours of CR~Boo and V803 Cen are comparable to those of MGAB-V270, ZTF18aaxuusk, and ASASSN-19ct, as can be seen in Figure~\ref{fig:CMD}, they exhibit brighter $M_\mathrm{G}$ magnitudes than these three systems. Similarly, ASASSN-14cc exhibits larger $M_\mathrm{G}$ magnitude comparable to those of CR~Boo and V803 Cen. While its $\mathrm{BP}-\mathrm{RP}$ colours are comparable to the studied systems, its difference between high and low state is larger than what we observed for ASASSN-19ct and V803 Cen. The position of KL Dra in the colour-magnitude diagram shows that it belongs among the reddest AM CVns. Its colour index is larger that those of MGAB-V270, ZTF18aaxuusk, or ASASSN-19ct and it is also brighter than these three systems.

\subsection{On the origin of standstill-like phases}

Several suggestions have been made to account for the rebrightenings that follow superoutbursts in CVs \citep[see][for a list of possible models]{2015PASJ...67..108K,2021A&A...650A.114H}. Here, we focus on the proposal by \cite{2021A&A...650A.114H}  who used the DIM to successfully model the light curves of two WZ Sge-type CVs showing rebrightenings. In their model, the mass transfer rate $\dot{M}_\mathrm{tr}$ from the donor increases during and after a superoutburst as a result of irradiation of the secondary star by radiation emitted by the heated accreting white dwarf. $\dot{M}_\mathrm{tr}$ returns to its pre-outburst value only after a cooling time of the white dwarf envelope. Echo outbursts are produced during this high mass-transfer phase, as cooling fronts are not able to propagate all the way down to the inner radius of the disc, but instead are reflected into heating fronts near the inner edge of the disc, because the mass accumulation time at high mass transfer rates and small radii becomes shorter than the viscous time what enables surface density to cross its critical value there,  in much the same way as was proposed by \cite{2000A&A...353..244H} for RZ LMi, an ER UMa system which shows very short supercycles (see their Figure 4). They also included additional ingredients, such as the truncation of the inner accretion disc at low mass transfer rates, or disc stream overflow. Moreover, because they were interested in WZ Sge stars, which have recurrence rates as long as decades, they assumed that the superoutburst was triggered by an enhancement of the mass-transfer rate in a system that is stable on the cold branch. 

A similar mechanism could explain the outburst behaviour observed in MGAB-V270, ZTF18aaxuusk and ASASSN-19ct. The same ingredients should be included, i.e. secondary irradiation leading to enhanced mass transfer and truncation of the inner accretion disc due to the magnetic field of the accretor at low mass transfer rates. The main difference is that superoutburst need not be triggered by a mass-transfer enhancement but can simply result from the regular increase of the disc mass (even though this can still be possible). The long duration of superoutburst would be explained by the increase of the mass-transfer rate in all cases.

The presence of standstills indicates that the mass transfer rate $\dot{M}_\mathrm{tr}$ becomes larger than the critical rate $\dot{M}^+_\mathrm{crit}(r_\mathrm{out})$ for the disc to stay on the hot branch at the outer edge because $\dot{M}^+_\mathrm{crit}\propto r^{2.65}$. This is true only in a steady state situation, which is not the case here since the mass distribution in the accretion disc at the end of an outburst or during the echo outburst phase is not that of a steady disc, but such an assumption is not terribly wrong and is a useful guide to understand the outburst physics. When the superoutburst stops, one naturally expects echo outbursts because the mass-transfer rate has to be less than (but still close to)  $\dot{M}^+_\mathrm{crit}(r_\mathrm{out})$; in favour of that interpretation is that for a disc with $r_\mathrm{out}\sim1.1\times10^{10}$ cm  composed of $98\%$ helium and $2\%$ metals $\dot{M}^+_\mathrm{crit}(r_\mathrm{out})=1.5\times10^{-9}\,\,\mathrm{M}_{\odot}/\mathrm{yr}$ (higher metallicity of the disc or/and disc irradiation may lower $\dot{M}^+_\mathrm{crit}(r_\mathrm{out})$ value by as much as order of magnitude) and according to evolutionary models of \citet{2005ApJ...624..934D} the intrinsic (with no enhancement) mass transfer rates from the He WD donor in systems with $P_\mathrm{orb}\sim 30$ min are in the range of $10^{-10}-10^{-9}\,\,\mathrm{M}_{\odot}/\mathrm{yr}$.
What may come as a surprise is the presence of a standstill \emph{after} the echo outburst phase, because one does not expect within the considered model that the mass transfer rate increases with time with no relation to an outburst. This could however happen, as a result for example of a decrease of the accretion disc opening angle and hence of the shielding of the secondary. It is also possible that $r_\mathrm{out}$ decreases, for example because the disc stream overflow is reduced as the mass transfer decreases; as $\dot{M}^+_\mathrm{crit}$ is a sensitive function of radius, the ratio $\dot{M}_\mathrm{tr}/\dot{M}^+_\mathrm{crit}(r_\mathrm{out})$ increases with time even though $\dot{M}_\mathrm{tr}$ decreases.

Trying to reproduce specific features of the observed light curves using the DIM might not lead to a better understanding of the detailed physics at play in these systems because there are far too many unconstrained free parameters, among which is the response of the secondary star to irradiation. It should also be stressed out that the 1D version of the DIM fails to account for the tidal distortion of the accretion disc, which is significant, as shown by the existence of superhumps. In other words, one may try to get a $\dot{M}_\mathrm{tr} (t)$ function that results in a light curve which matches observations -- and such a function probably exists -- but this would not be a proof that the ingredients of the model are correct.

The above scenario could be used to explain also the superoutbursts and its rebrightenings of CR Boo and V803 Cen. These two systems, however, have much shorter supercycles and they experience long-lasting standstills, which suggests that they have mass transfer rates that are closer to the critical $\dot{M}_\mathrm{crit}^+$ than is the case of MGAB-V270, ZTF18aaxuusk, or ASASSN-19ct. While ASASSN-14cc recurrence time is comparable with that of MGAB-V270, the fact that its superoutbursts are only occasionally followed by rebrightenings suggests that the properties of this system differs from those of our three studied targets. The diverse morphology of superoutbursts and following rebrightenings observed in KL Dra shows that while the described scenario could apply to this system to some extent, it is likely more complex. The mean superoutburst recurrence time of KL Dra was determined by \cite{2026PASA...43...70S} as $T_\mathrm{SO}=60.4\,\mathrm{d}$ is also shorter than the ones determined for our three studied systems, showing that its mass transfer rate is closer to its $\dot{M}_\mathrm{crit}^+$ value, as for CR Boo and V803 Cen.

\section{Conclusions}

We analysed available photometry of three AM~CVn systems showing similar peculiar behaviour following their superoutbursts. All three systems have almost identical superhump periods and their spectra exhibit similar spectral features. Two of these systems, MGAB-V270 and ZTF18aaxuusk, show standstill-like phases during their post-superoutburst brightness variations. In the case of MGAB-V270, we detected a superhump period during these phases. The third system, ASASSN-19ct, does not show a standstill-like phase, but it shows an increasing trend of the minimum brightness between echo outbursts, which is a behaviour detected also in MGAB-V270 and ZTF18aaxuusk prior to their standstill-like phases. 

These standstill-like phases are connected to the post-superoutburst activity of the system and they are shorter than the typical standstill observed in Z Cam-type systems. This suggests that they are a distinct phenomenon.

The comparison with other AM~CVns showed that a similar behaviour could also exist in other systems; however, the presence of standstill-like phases cannot be unambiguously confirmed due to the limited coverage of available observations. While MGAB-V270, ZTF18aaxuusk, and ASASSN-19ct have very similar characteristics, i.e. almost identical orbital period or comparable recurrence times of superoutbursts, the presence of similar outbursting behaviour in the comparison systems suggests that this type of outburst activity may occur in AM~CVn systems spanning a wide range of properties. 

We suggest that these properties could be accounted for by variations of the mass-transfer rate during and after a superoutburst as a result of irradiation of the secondary star by the heated accreting white dwarf and by the accretion disc. We caution, however, that the large number of poorly constrained free parameters severely limits the predictive power of such models, even if they were to successfully reproduce the observed light curves.

\section*{Acknowledgment}
JK, LRS and RJO acknowledge support from NASA grants NNH22ZDA001N-6152 and 80NSSC24K0638.
We Thank J. van Roestell for providing us with spectroscopic observations of MGAB-V270.
We gratefully acknowledge the contributions of the AAVSO observer community, whose photometric data and metadata resources were used in this study and made available through the AAVSO’s scientific archives. 
This work has made use of data from the European Space Agency (ESA) mission
{\it Gaia} (\url{https://www.cosmos.esa.int/gaia}), processed by the {\it Gaia}
Data Processing and Analysis Consortium (DPAC,
\url{https://www.cosmos.esa.int/web/gaia/dpac/consortium}). Funding for the DPAC
has been provided by national institutions, in particular the institutions
participating in the {\it Gaia} Multilateral Agreement.
This work made use of Astropy: \footnote{https://www.astropy.org} a community-developed core Python package and an ecosystem of tools and resources for astronomy \citep{astropy:2013, astropy:2018, astropy:2022}.
This research has made use of the SIMBAD database, CDS, Strasbourg Astronomical Observatory, France

\section*{Data Availability}

This research is based on publicly available observations and previously published data.

\bibliographystyle{aa_link}
\bibliography{References.bib}

\appendix
\renewcommand{\thefigure}{\thesection.\arabic{figure}}
\renewcommand{\thetable}{\thesection.\arabic{table}}

\begin{table*}[]
\section{List of TESS light curves}
\caption{Table listing all TESS light curves used in this work and outbursting activity found in the light curves. NO stands for normal outburst, SO stands for superoutburst, EO stands for echo outburst. Light curves strongly affected by TESS instrumental effects are labelled as TIE.}
\begin{tabular}{lccl | lccl}
\toprule
Name         & \multicolumn{1}{l}{Sector} & Cadence & Comments                                         & Name         & \multicolumn{1}{l}{Sector} & Cadence & Comments               \\
\hline
MGAB-V270    & 16                         & 30 min  &                                                  & ZTF18aaxuusk & 57                         & 200 s   &                        \\
             & 17                         & 30 min  &                                                  &              & 58                         & 200 s   &                        \\
             & 24                         & 30 min  &                                                  &              & 59                         & 200 s   &                        \\
             & 56                         & 120 s   &                                                  &              & 60                         & 200 s   &                        \\
             & 57                         & 120 s   &                                                  &              & 73                         & 120 s   &                        \\
             & 58                         & 120 s   & 1 SO, 7 EOs                                      &              & 74                         & 120 s   &                        \\
             & 76                         & 200 s   & 1 SO, 7 EOs, standstill, 2 NOs                   &              & 76                         & 120 s   &                        \\
             & 77                         & 200 s   &                                                  &              & 77                         & 120 s   &                        \\
             & 78                         & 200 s   &                                                  &              & 79                         & 120 s   &                        \\
             & 83                         & 200 s   &                                                  &              & 80                         & 200 s   &                        \\
             & 84                         & 200 s   & 1 SO, 9 EOs, stanstill, 1 NO                     &              & 81                         & 200 s   &                        \\
             & 85                         & 200 s   &                                                  &              & 82                         & 120 s   &                        \\
\cline{1-4}
ZTF18aaxuusk & 14                         & 30 min  & 1 SO, 9 EOs                                      &              & 83                         & 120 s   & 1 SO, 6 EOs            \\
             & 15                         & 30 min  & 2 EOs, 1 NO                                      &              & 84                         & 200 s   &                        \\
             & 16                         & 30 min  &                                                  &              & 85                         & 200 s   &                        \\
             & 17                         & 30 min  &                                                  &              & 86                         & 200 s   &                        \\
\cline{5-8}
             & 18                         & 30 min  &                                                  & ASASSN-19ct  & 10                         & 30 min  & 1 SO, 18 EOs           \\
             & 19                         & 30 min  &                                                  &              & 36                         & 120 s   &                        \\
             & 20                         & 30 min  &                                                  &              & 37                         & 120 s   & 1 NO, 1 SO             \\
             & 21                         & 30 min  &                                                  &              & 63                         & 120 s   & 1 NO                   \\
             & 22                         & 30 min  &                                                  &              & 90                         & 120 s   &                        \\
             & 23                         & 30 min  & 1 SO, 6 EOs                                      &              & 99                         & 200 s   &                        \\
             & 24                         & 30 min  & 4 EOs, standstill, 1 NO                          &              & 100                        & 200 s   &                        \\
             & 25                         & 30 min  &                                                  &              & 101                        & 120 s   &                        \\
\cline{5-8}
             & 26                         & 30 min  &                                                  & V803 Cen     & 11                         & 30 min  & 5 NOs                  \\
             & 40                         & 10 min  & 9 EOs                                            &              & 37                         & 120 s  & 2 NOs                  \\
             & 41                         & 10 min  &                                                  &              & 64                         & 120 s   & 1 NO, 1 SO, 12 EOs     \\
             & 47                         & 10 min  & 1 SO, 10 EOs                                     &              & 101                        & 120 s   & 22 EOs, 1 SO           \\
             & 48                         & 10 min  & 2 EOs                                            &              & 102                        & 200 s   &                        \\
\cline{5-8}
             & 49                         & 10 min  &                                                  & ASASSN-14cc  & 1                          & 30 min  & 2 NOs, 1 stunted, 1 SO \\
             & 50                         & 10 min  &                                                  &              & 27                         & 10 min  & 2 NOs                  \\
             & 51                         & 10 min  &                                                  &              & 28                         & 10 min  & 1 SO, TIE              \\
             & 52                         & 10 min  &                                                  &              & 102                        & 120 s   & 1 NO                   \\
             & 53                         & 10 min  &                                                  &              & 103                        & 120 s   & 1 SO                   \\
             & 54                         & 10 min  & 1 SO                                             &              & 104                        & 200 s   & 1 NO, 1 SO             \\
             & 55                         & 10 min  & EOs, TIE &              & 105                        & 200 s   &                        \\
             & 56                         & 200 s   &                                                  &              & \multicolumn{1}{l}{}       &         &              
\botrule
\end{tabular}
\label{tab:TESS_ALL}
\end{table*}

\FloatBarrier

\begin{figure*}
\section{TESS light curves}
    \centering
    \includegraphics[width=0.98\linewidth]{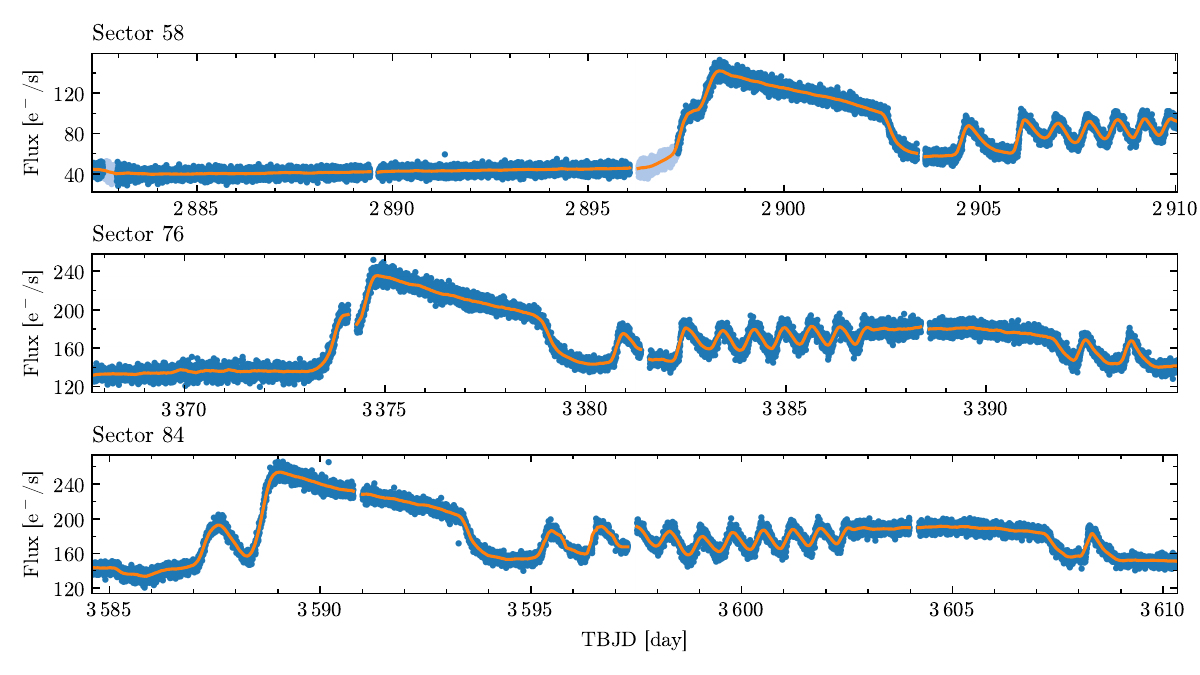}

    \caption{TESS light curves of MGAB-V270, only sectors exhibiting outbursting activity are shown.}
    \label{fig:TESS_ALL_MGAB-V270}
\end{figure*}

\begin{figure*}
    \centering
    \includegraphics[width=0.98\linewidth]{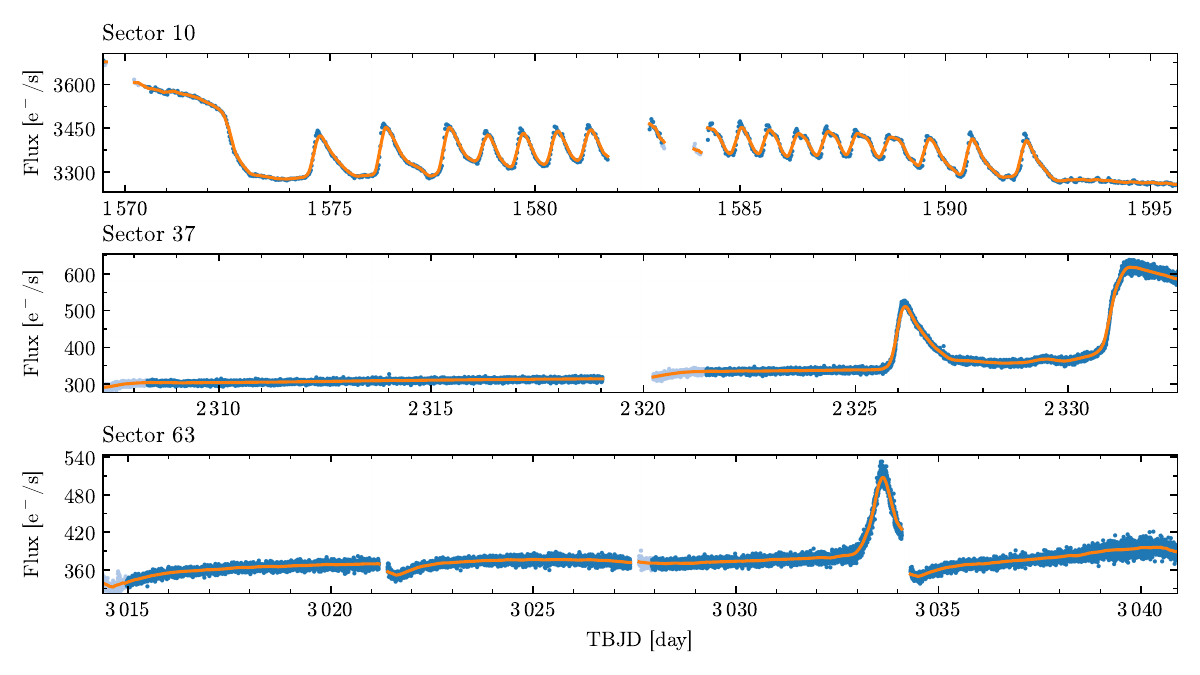}

    \caption{TESS light curves of ASASSN-19ct, only sectors exhibiting outbursting activity are shown.}
    \label{fig:TESS_ALL_ASASSN-19ct}
\end{figure*}

\begin{figure*}
    \centering
    \includegraphics[width=0.98\linewidth]{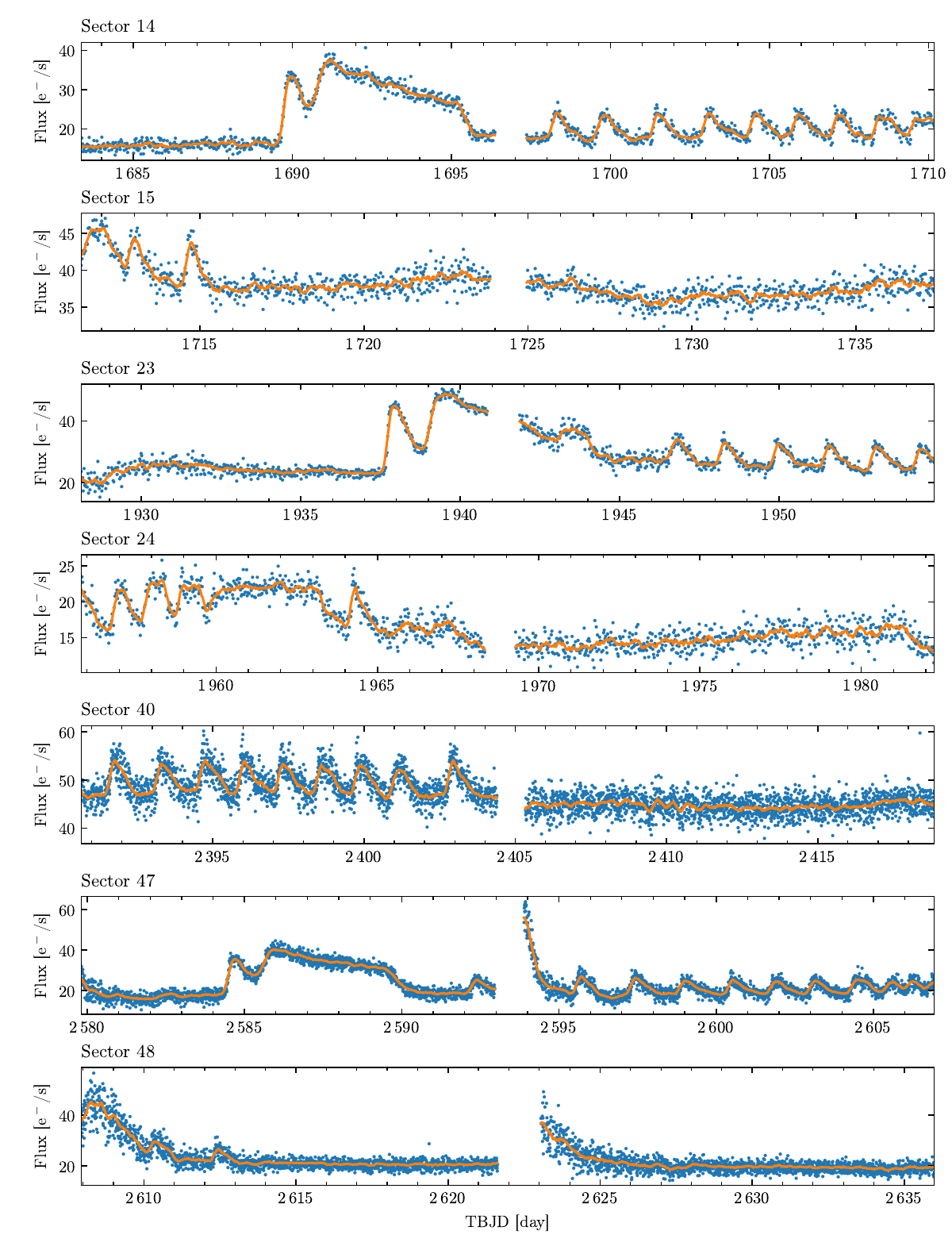}

    \caption{TESS light curves of ZTF18aaxuusk, part I. Only sectors exhibiting outbursting activity are shown. Note that light curves of Sectors 47 and 48 are affected by instrumental artefacts.}
    \label{fig:TESS_ALL_ZTF18aaxuusk_a}
\end{figure*}

\begin{figure*}
    \centering
    \includegraphics[width=0.98\linewidth]{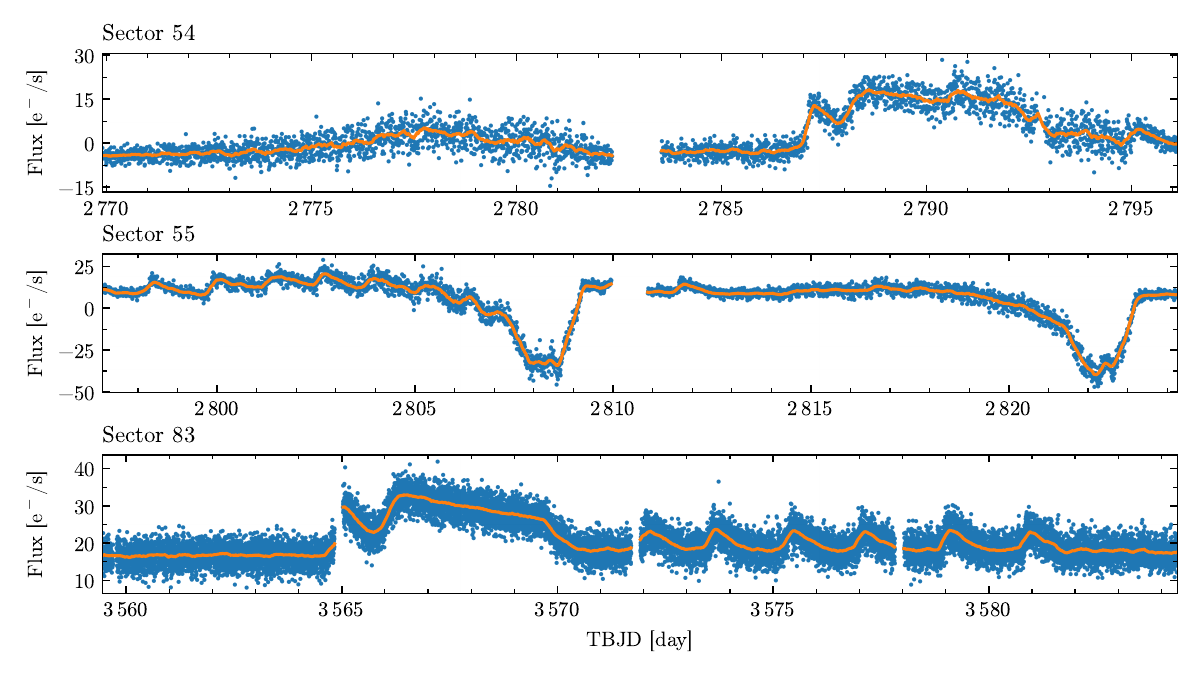}

    \caption{TESS light curves of ZTF18aaxuusk, part II. Only sectors exhibiting outbursting activity are shown.  Note that light curves of Sectors 54 and 55 are affected by instrumental artefacts.}
    \label{fig:TESS_ALL_ZTF18aaxuusk_b}
\end{figure*}

\begin{figure*}
    \centering
    \includegraphics[width=0.98\linewidth]{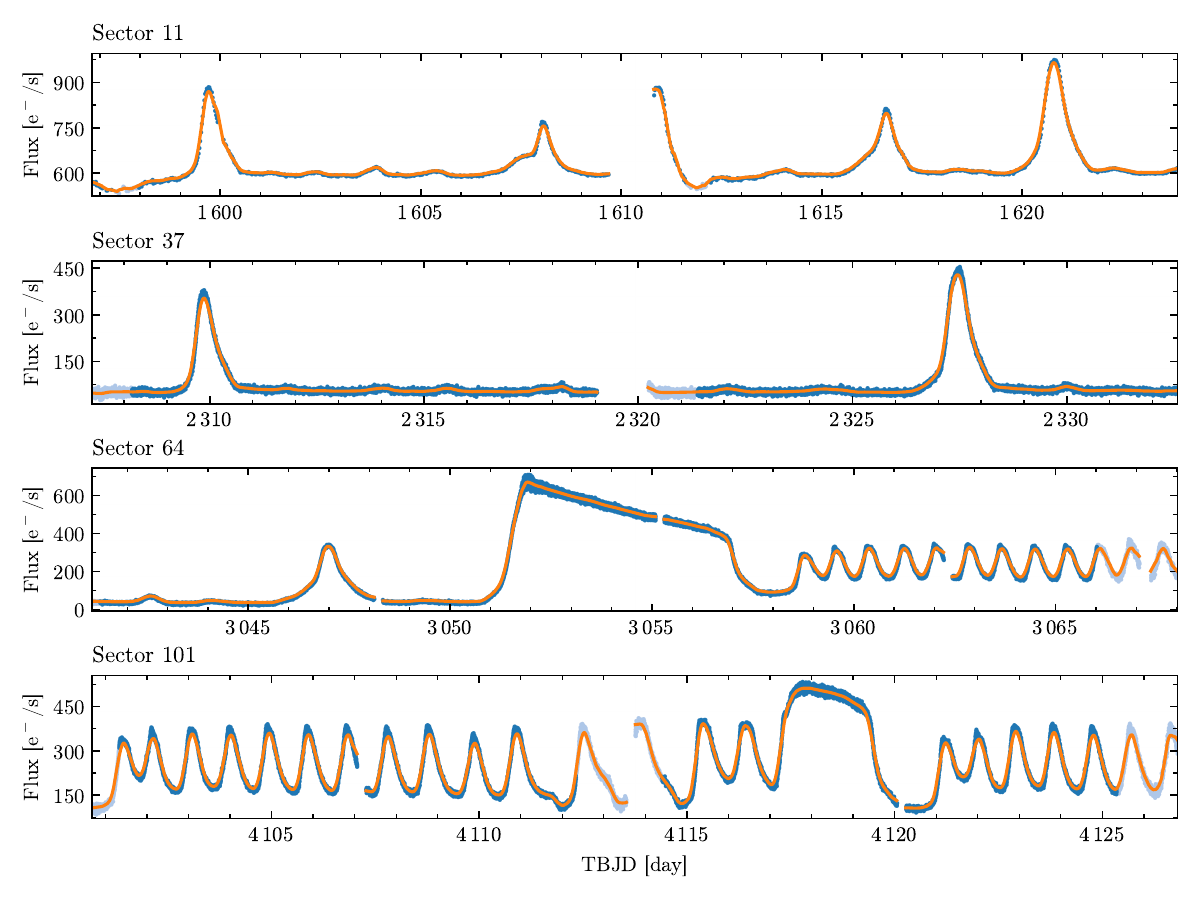}

    \caption{TESS light curves of V803 Cen, only sectors exhibiting outbursting activity are shown. Note that the light curve from Sector 101 was obtained during a standstill.}
    \label{fig:TESS_ALL_V803Cen}
\end{figure*}

\begin{figure*}
    \centering
    \includegraphics[width=0.98\linewidth]{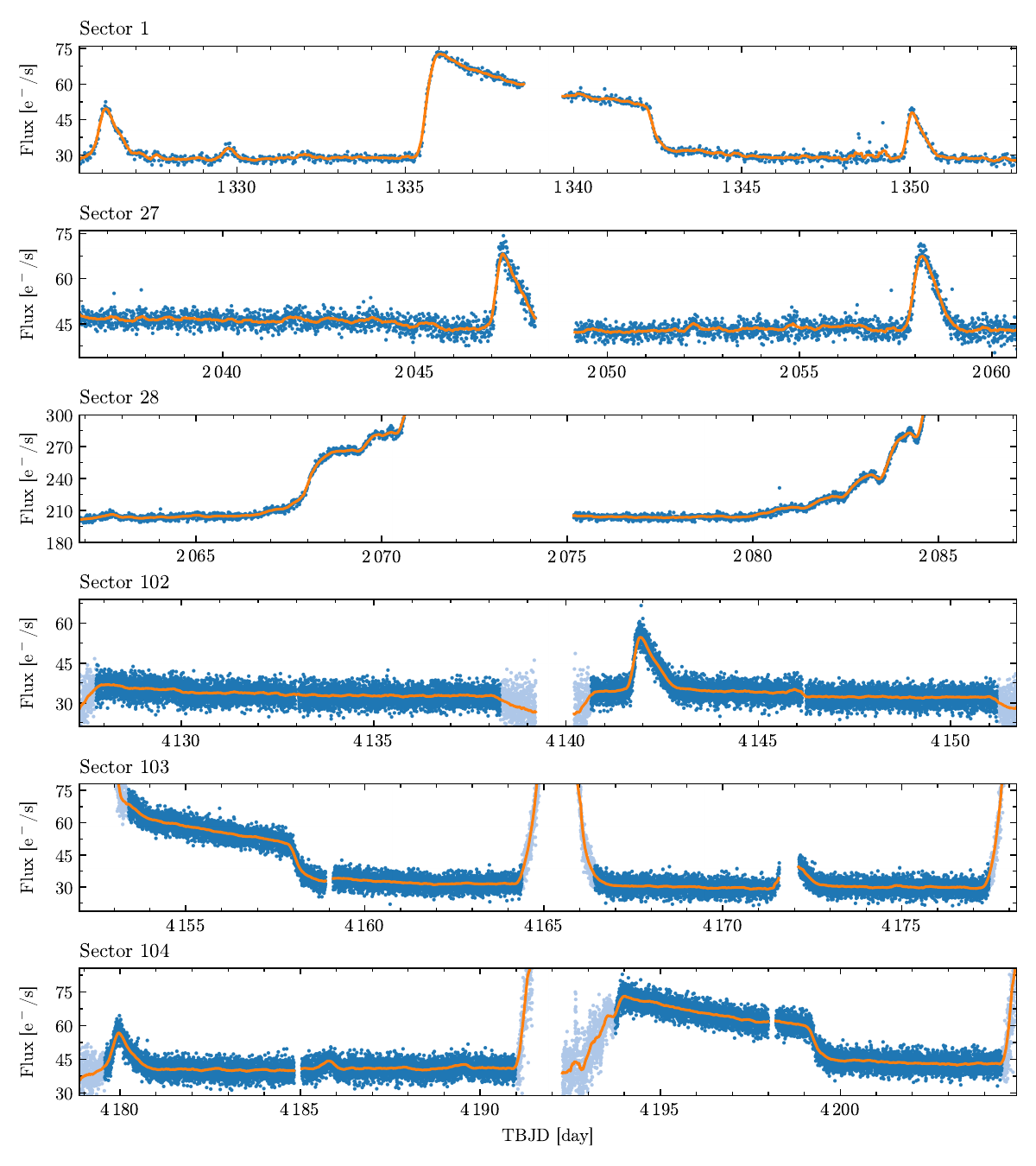}

    \caption{TESS light curves of ASASSN-14cc, only sectors exhibiting outbursting activity are shown. Note that Sector 28 light curve is heavily affected by instrumental artefacts.}
    \label{fig:TESS_ALL_ASASSN-14cc}
\end{figure*}

\FloatBarrier

\begin{figure*}
\section{Long-term photometry}
    \centering
    \includegraphics[width=0.98\linewidth]{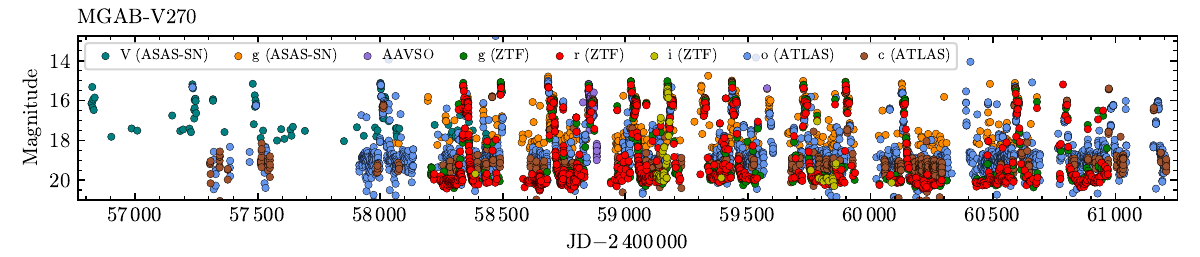}

    \includegraphics[width=0.98\linewidth]{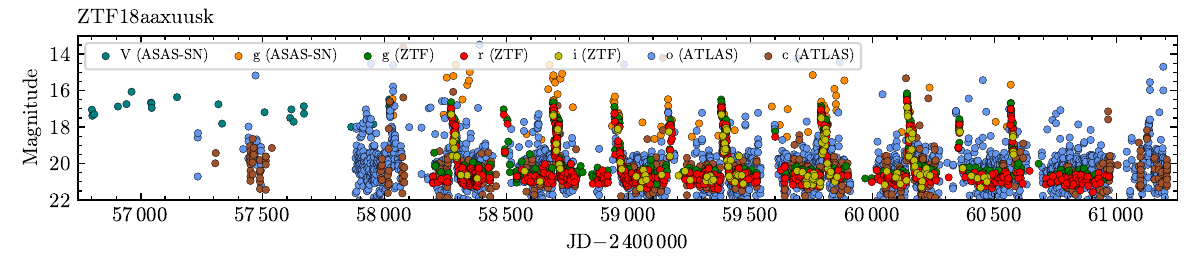}

    \includegraphics[width=0.98\linewidth]{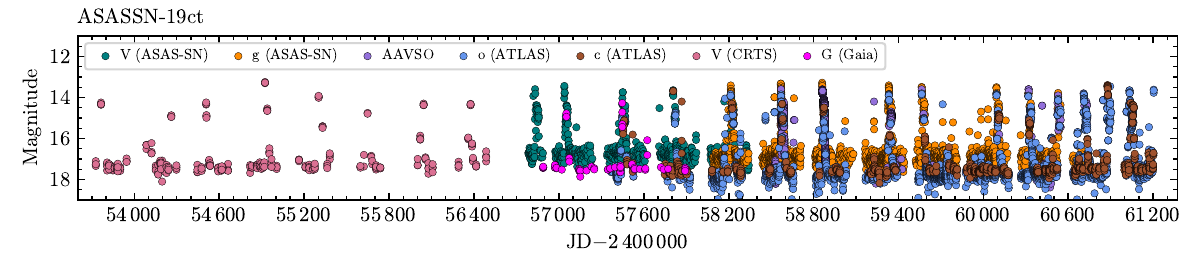}

    \includegraphics[width=0.98\linewidth]{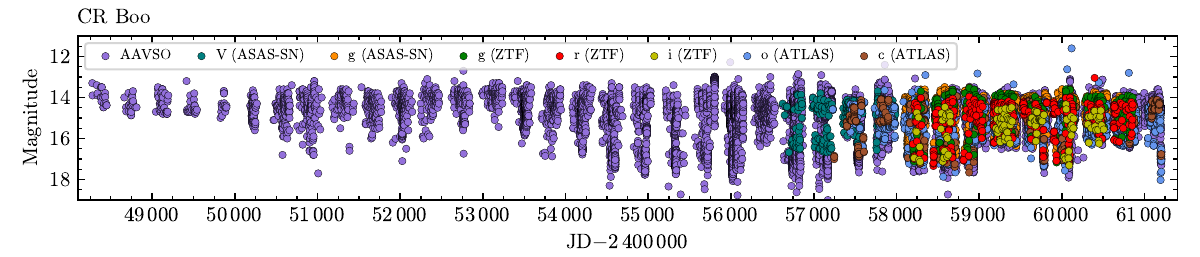}

    \includegraphics[width=0.98\linewidth]{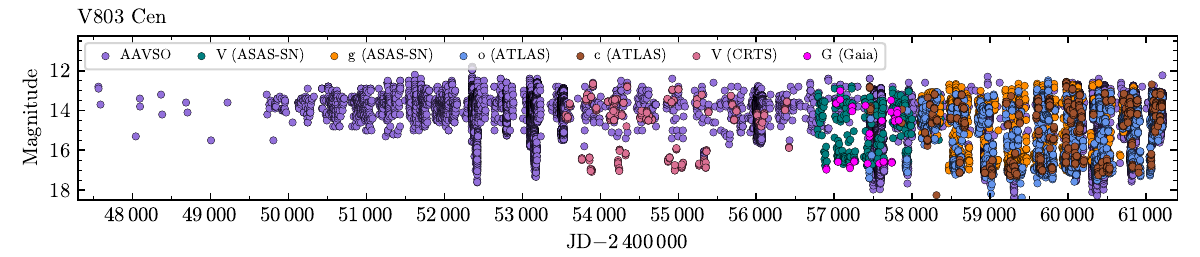}

    \includegraphics[width=0.98\linewidth]{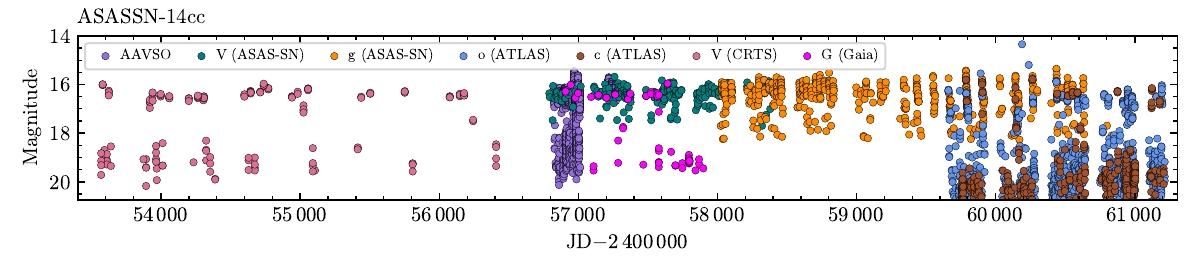}

    \caption{Long-term photometry of the studied targets.}
    \label{F:LTLC}
\end{figure*}

\FloatBarrier

\begin{figure*}
\section{Comparison of long-term light curves and TESS photometry }
    \centering
    \includegraphics[width=0.98\linewidth]{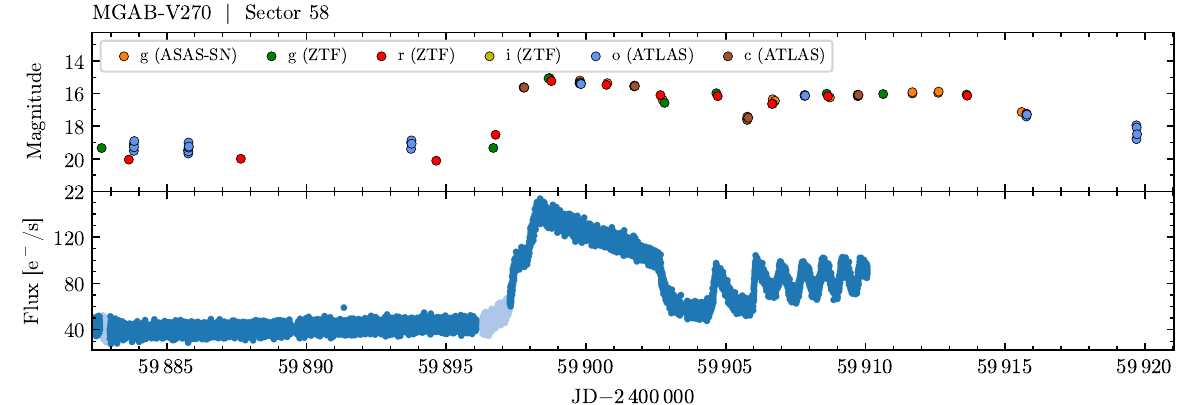}

    \includegraphics[width=0.98\linewidth]{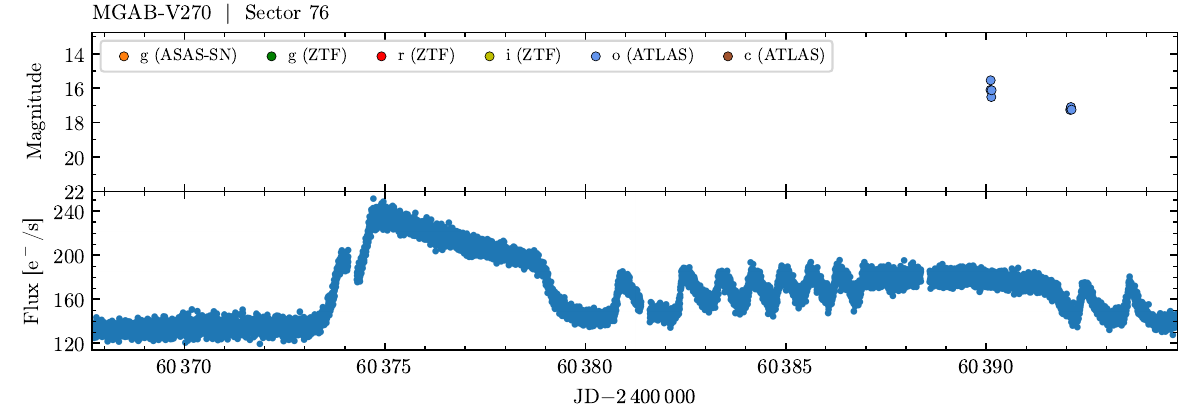}

    \includegraphics[width=0.98\linewidth]{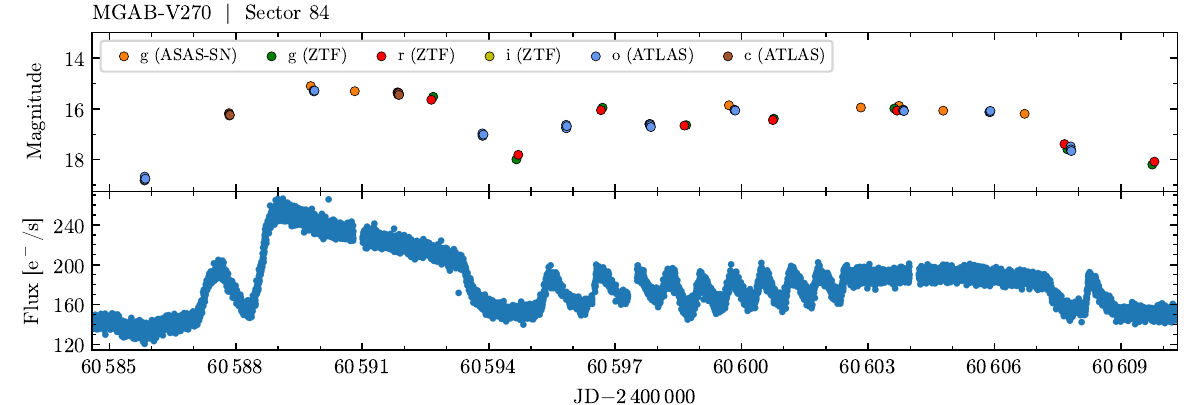}

    \caption{Comparison to ground-based photometry from various surveys (top panels) and TESS photometry (bottom panels) of MGAB-V270. TESS data with low-quality flags are plotted in light colour. Note the different time scales of the presented light curves which were chosen so that they cover the full TESS light curve and the full outbursting activity.}
    \label{F:MGAB_COMP_AP}
\end{figure*}

\begin{figure*}
    \centering
    \includegraphics[width=0.98\linewidth]{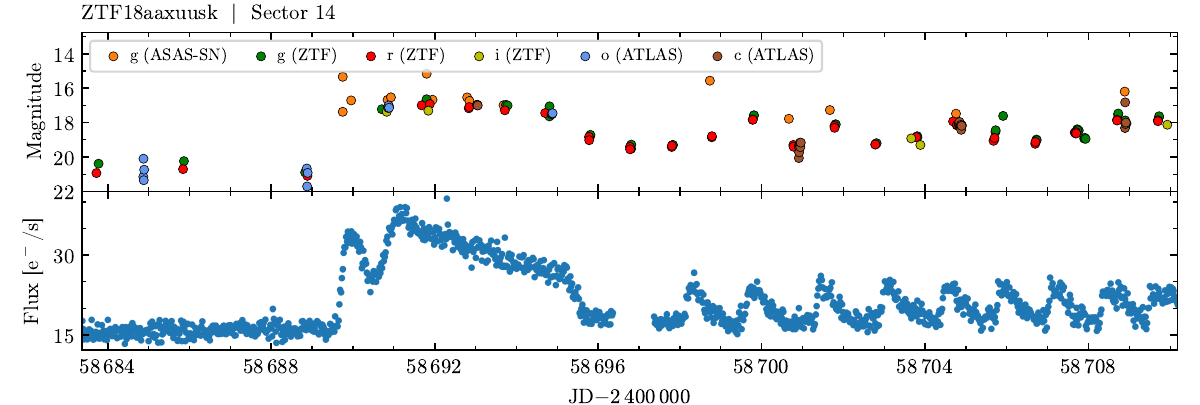}

    \includegraphics[width=0.98\linewidth]{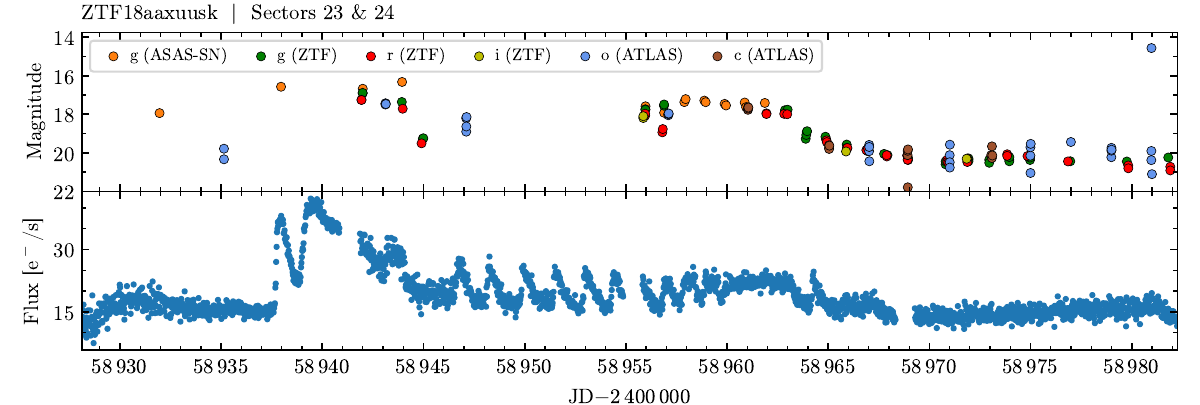}

    \includegraphics[width=0.98\linewidth]{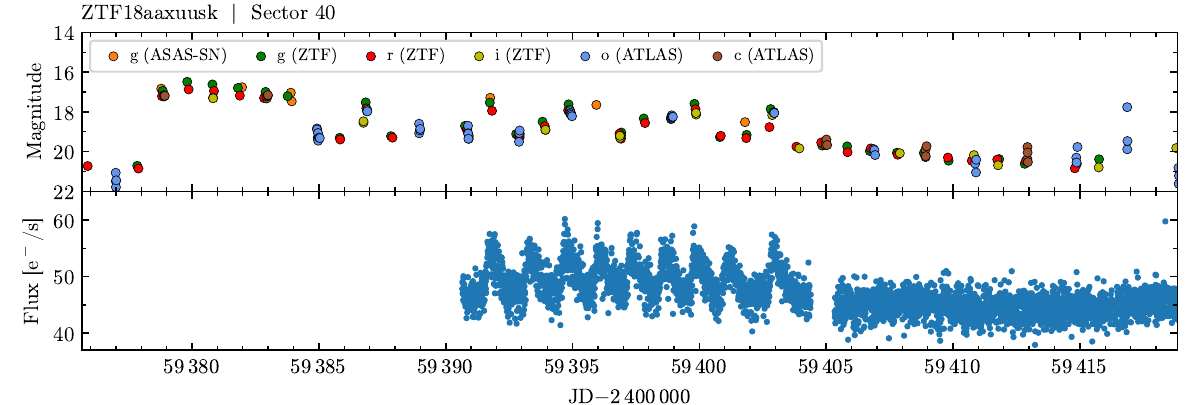}

    \caption{Comparison of ground-based photometry from various surveys (top panels) and TESS photometry (bottom panels) of ZTF18aaxuusk. The flux of Sector 23 is shifted such that its quiescent level matches the quiescent flux observed in Sector 24. 
    Note the different time scales of the presented light curves which were chosen so that they cover the full TESS light curve and the full outbursting activity.}
    \label{F:ZTF_COMP_AP}
\end{figure*}

\begin{figure*}[tb]
    \includegraphics[width=0.99\linewidth]{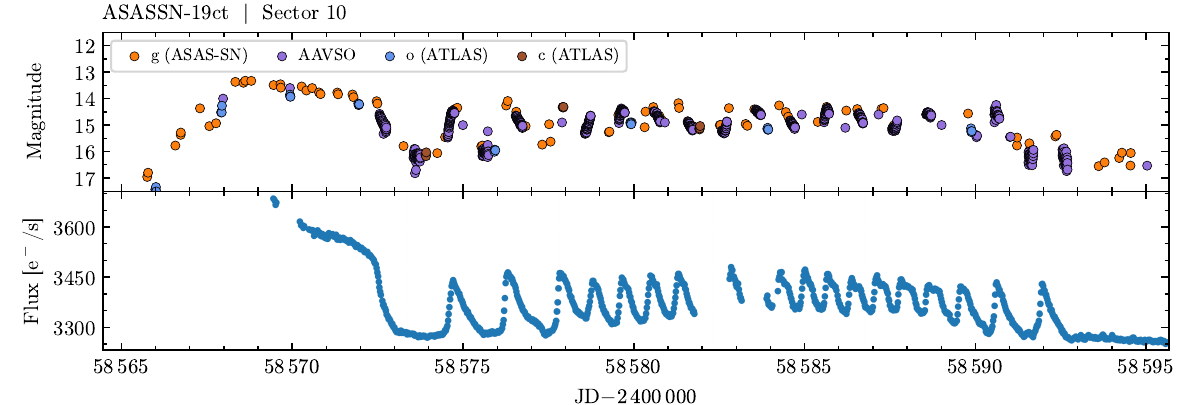}
    
    \includegraphics[width=0.99\linewidth]{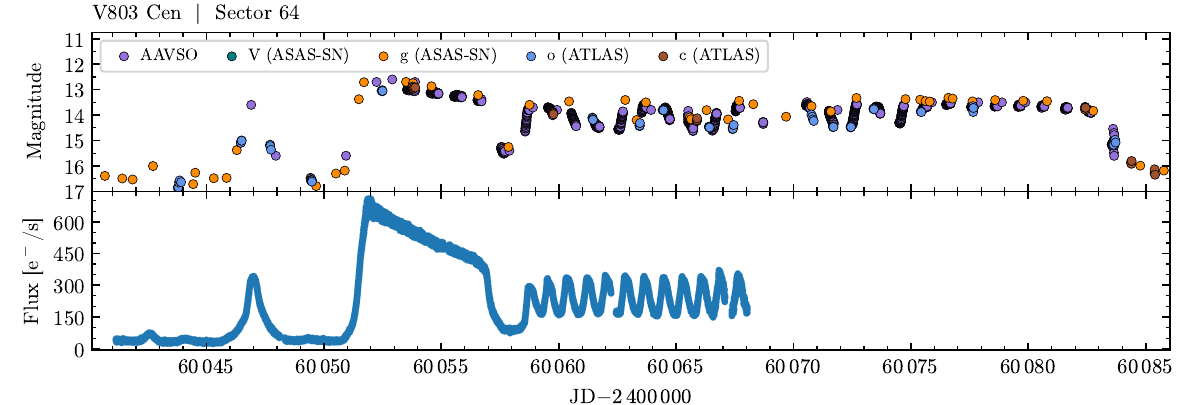}
    
    \includegraphics[width=0.99\linewidth]{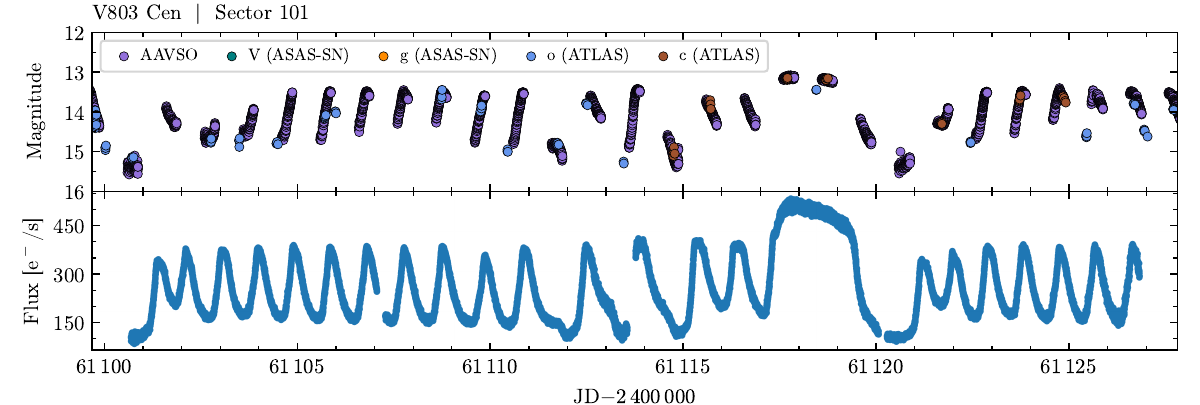}

    \caption{Comparison of ground-based photometry from various surveys (top panels) and TESS photometry (bottom panels) of ASASSN-19ct and V803 Cen. Note that the light curve of V803 Cen from Sector 101 was obtained during a standstill.}
    \label{fig:19ct_V803_COMP_AP}
\end{figure*}

\FloatBarrier

\begin{figure*}
\section{Two-dimensional periodograms of TESS light curves}
    \centering
    \includegraphics[width=0.49\linewidth]{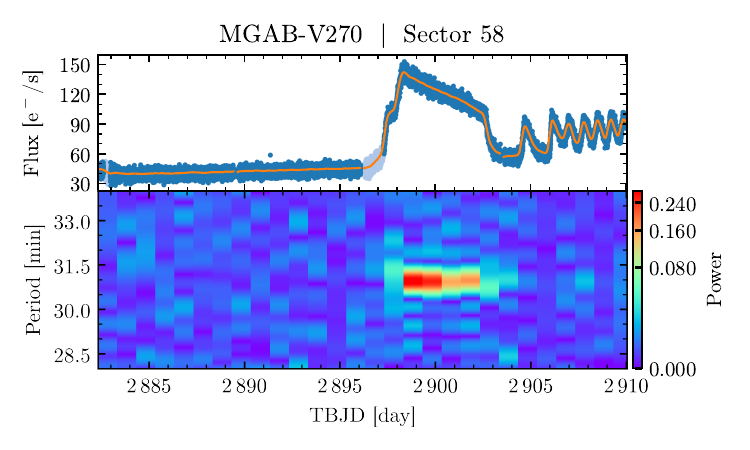}
    \includegraphics[width=0.49\linewidth]{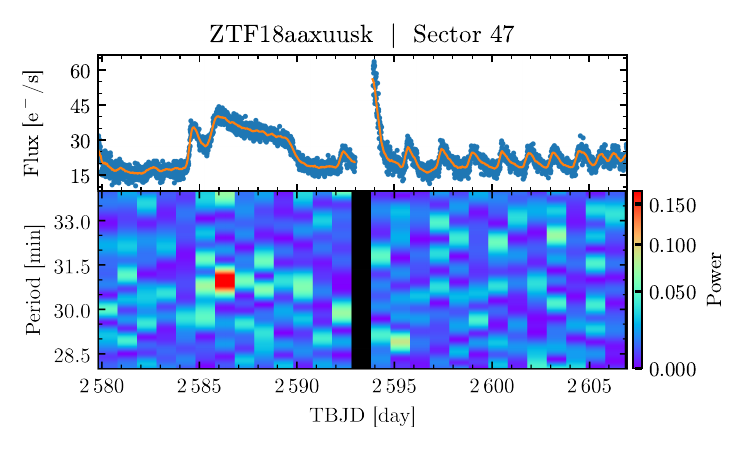}
    
    \includegraphics[width=0.49\linewidth]{MGAB-V270_sector76_2D_short.pdf}
    \includegraphics[width=0.49\linewidth]{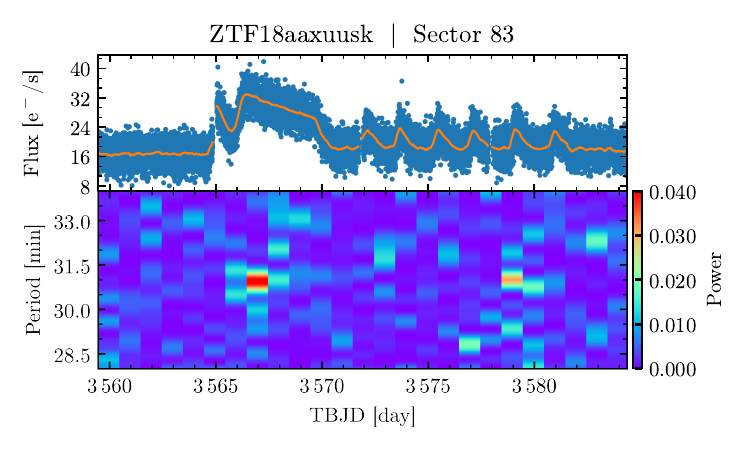}

    \includegraphics[width=0.49\linewidth]{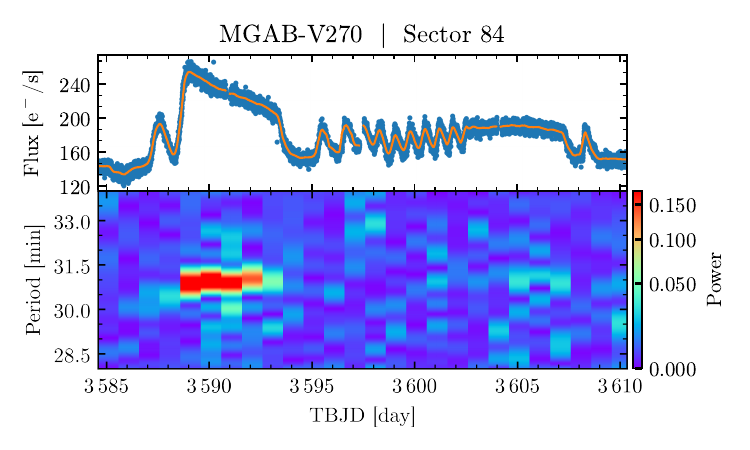}
    \includegraphics[width=0.49\linewidth]{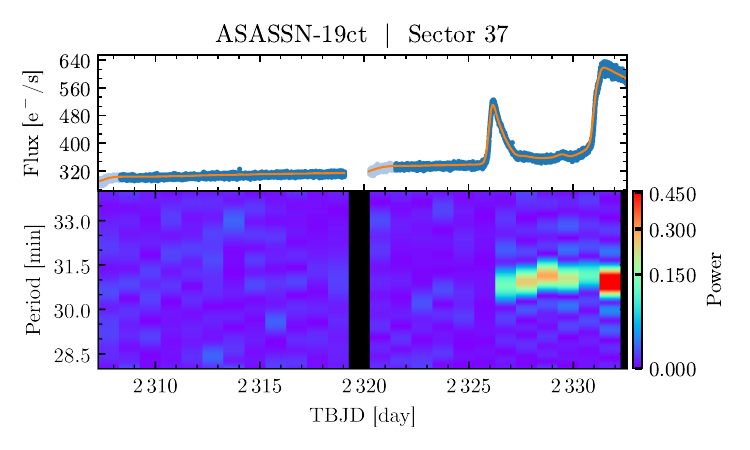}

    \caption{Two-dimensional Lomb-Scargle periodograms of TESS light curves showcasing outbursting activity of MGAB V270, ZTF18aaxuusk, and ASASSN-19ct. Top panels show the TESS photometry (blue points) and average light curves (orange lines). Photometry with bad quality flags are plotted in light-blue colour. Bottom panels show two-dimensional Lomb-Scargle periodograms based on the presented light curves.}
    \label{fig:2D_LomScargle_all}
\end{figure*}

\begin{figure*}
    \begin{minipage}[t]{.49\linewidth}\vspace{0pt}
        \includegraphics[width=\linewidth]{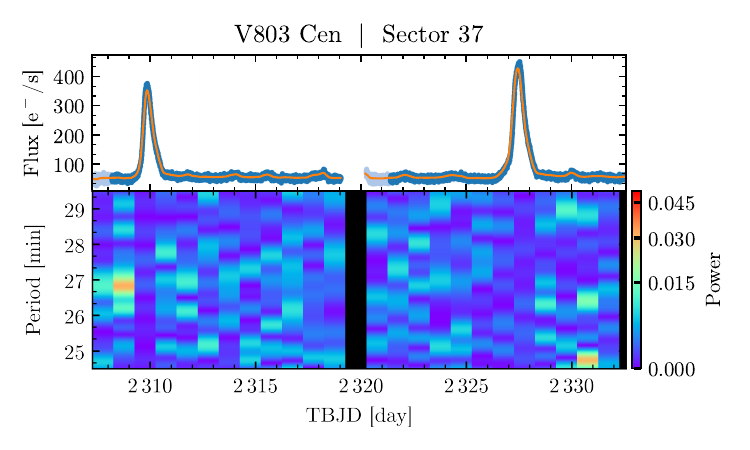}
        \includegraphics[width=\linewidth]{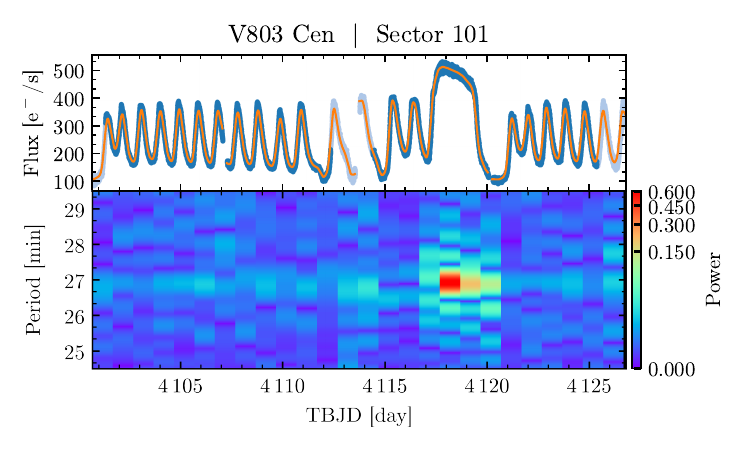}
    \end{minipage}
    \begin{minipage}[t]{.49\linewidth}\vspace{0pt}\raggedright
        \includegraphics[width=\linewidth]{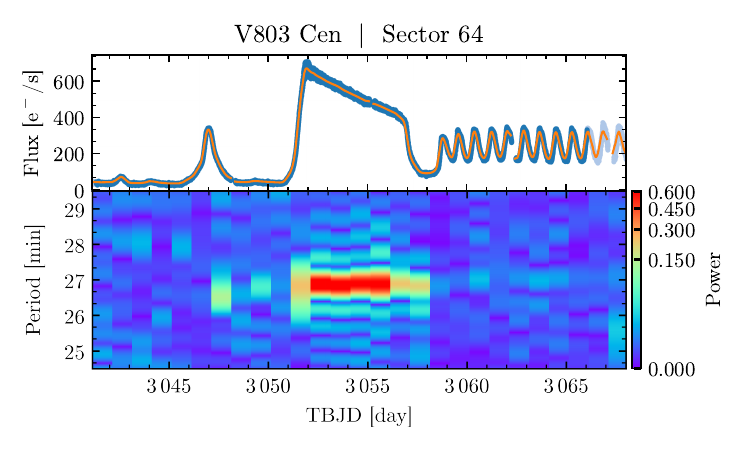}
    \end{minipage}

    \caption{Two-dimensional Lomb-Scargle periodograms of TESS light curves showcasing outbursting activity of V803 Cen. See Figure~\ref{fig:2D_LomScargle_all} for detailed description.}
    \label{fig:2D_LomScargle_V803}
\end{figure*}

\FloatBarrier

\begin{figure*}
\section{Lomb-Scargle periodograms of specific sections of TESS light curves}
 \begin{minipage}[t]{.49\linewidth}\vspace{0pt}
   \includegraphics[width=\linewidth]{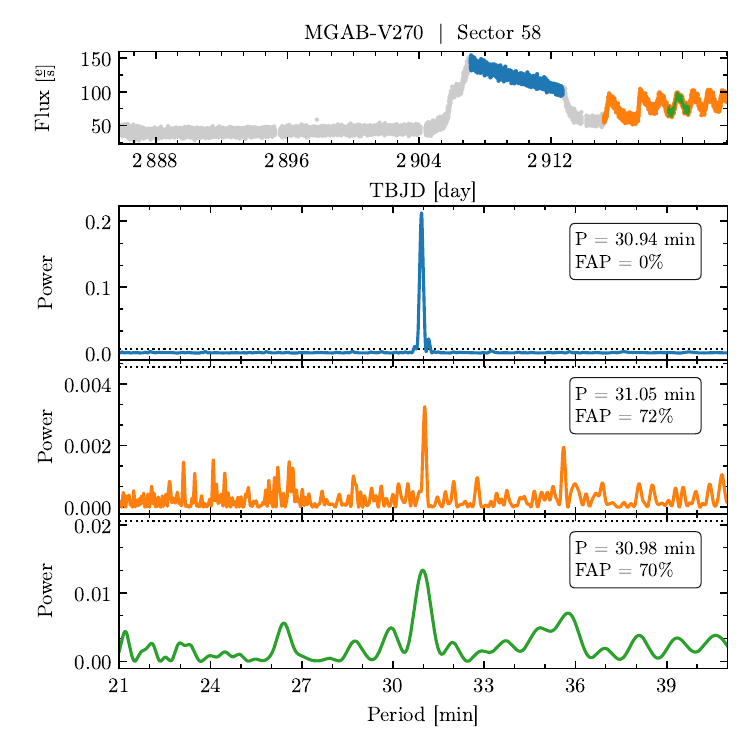}

   \includegraphics[width=\linewidth]{MGAB-V270_sector76_1D.pdf}
 \end{minipage}
 \begin{minipage}[t]{.49\linewidth}\vspace{0pt}\raggedright
   \includegraphics[width=\linewidth]{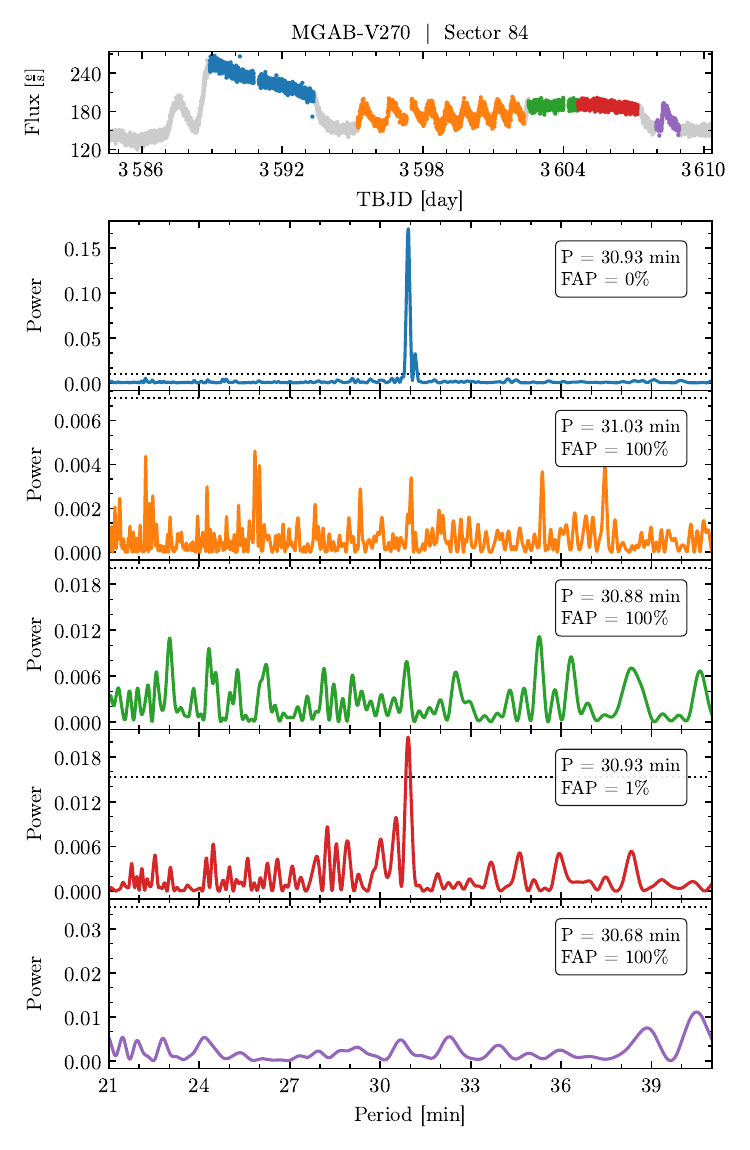}
   
 \end{minipage}
 
 \caption{Lomb-Scargle periodograms of different activity phases of MGAB-V270 based on TESS light curve from Sectors 58, 76, and 84.
    The orange light curve of Sector 84 covers the whole echo outburst phase, the green light curve focuses on the interval showing periodic variation in the two-dimensional periodogram presented in Figure~\ref{fig:2D_LomScargle_all}.
    The dotted line marks the power level corresponding to a $10\%$ FAP, the period and its FAP value of the strongest peak  within $30\,\mathrm{s}$ of the superhump period is presented for each periodogram.
    Note that both orange and green sections of the light curves in Sectors 76 and 84 contain a short observational gap.}
    \label{fig:MGABV270_1dLS_all}
\end{figure*}

\begin{figure*}
    \centering
    \includegraphics[width=0.49\linewidth]{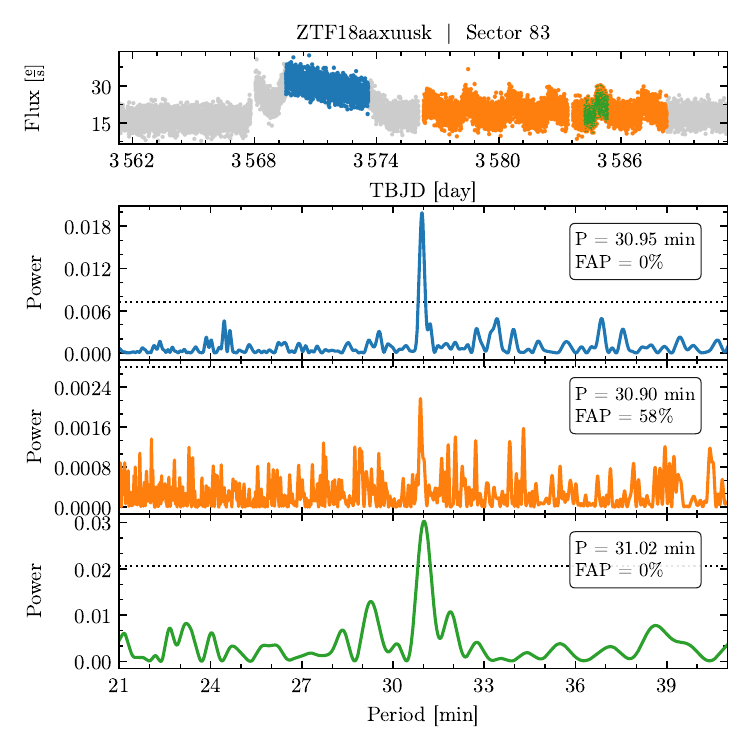}
    \includegraphics[width=0.49\linewidth]{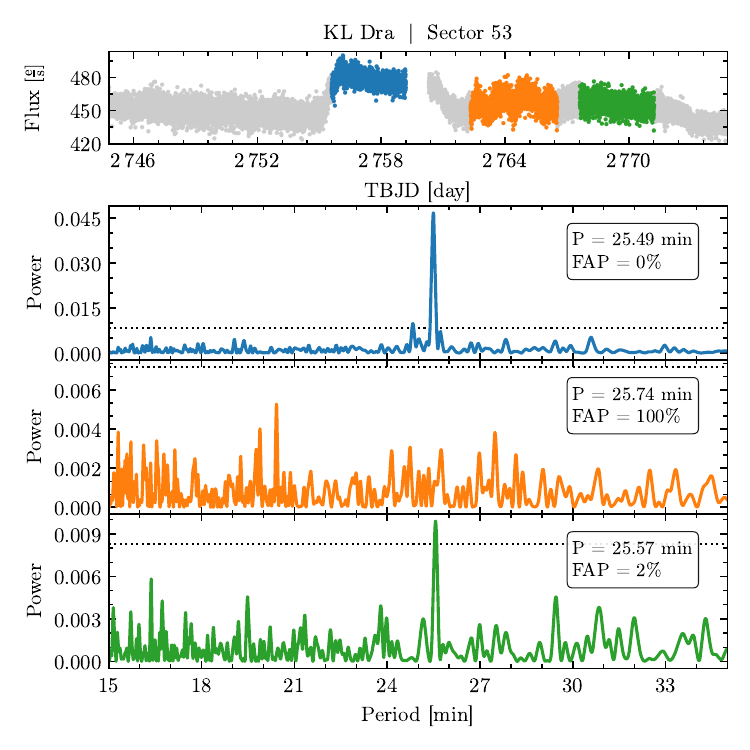}

    \includegraphics[width=0.49\linewidth]{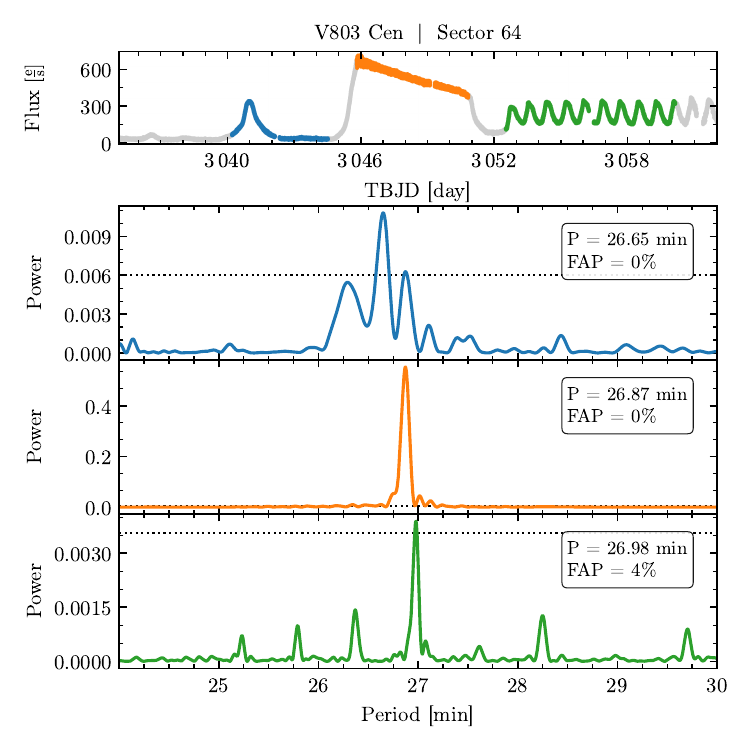}
    \includegraphics[width=0.49\linewidth]{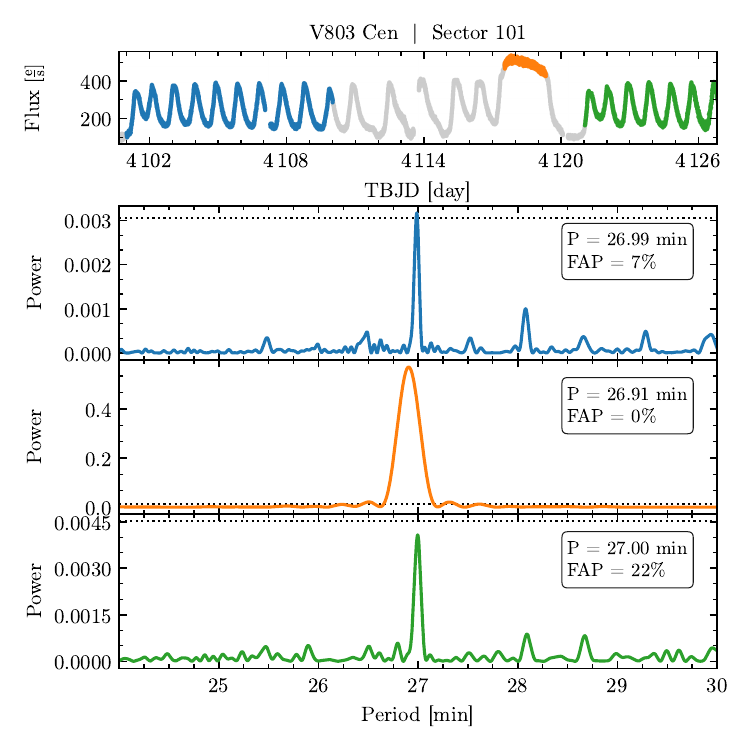}

    \includegraphics[width=0.49\linewidth]{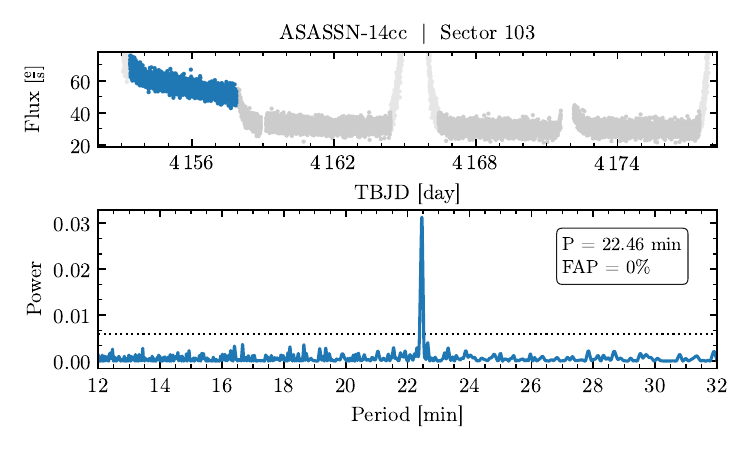}
    \includegraphics[width=0.49\linewidth]{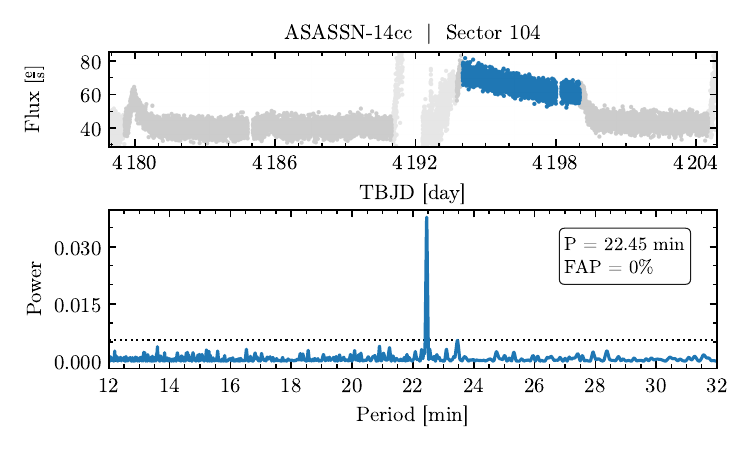}
    \caption{Lomb-Scargle periodograms of different activity phases of ZTF18aaxuusk, KL Dra, V803 Cen, and ASASSN-14cc based on TESS light curves. 
    The dotted line marks the power level corresponding to a $10\%$ FAP, the value of FAP of the strongest peak is within $30\,\mathrm{s}$ of the superhump period is presented for each periodogram.
    }
    \label{fig:ztf_803_kl_1dLS_14cc}
\end{figure*}

\end{document}